\documentclass[aps,pr,twocolumn,superscriptaddress,showpacs,preprintnumbers,amsmath,amssymb,fleqn]{revtex4}
\usepackage{graphicx} % Include figure files
\usepackage{dcolumn}  % Align table columns on decimal point

\graphicspath{{ps}}

\begin{document}

\vspace*{-3\baselineskip}
\resizebox{!}{3cm}{\includegraphics{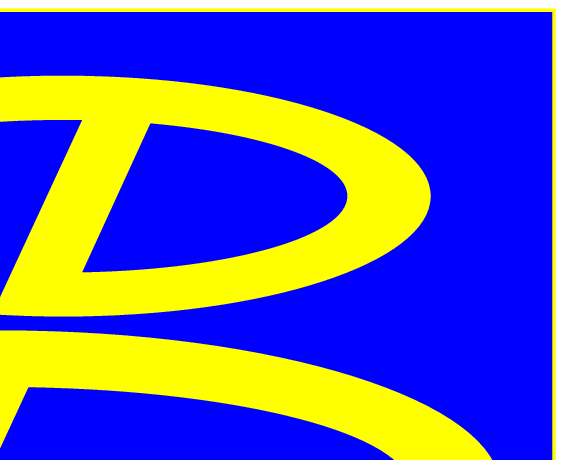}}

\preprint{\vbox{ \hbox{   }
			\hbox{Belle Preprint 2013-29}
                        \hbox{KEK Preprint 2013-57}
}}

\title{ \quad\\[1.0cm] Search for doubly charmed baryons and study of charmed strange baryons at Belle}

%%%% >>>>> insert the authorlist here. BEFORE the abstract !!!!! <<<<<
%%%% >>>>> from the authorship confirmation web page
%%% Name the file author.tex and use \input{author} to insert into your latex file.
\affiliation{University of the Basque Country UPV/EHU, 48080 Bilbao}
\affiliation{Beihang University, Beijing 100191}
%%%\affiliation{University of Bonn, 53115 Bonn}
\affiliation{Budker Institute of Nuclear Physics SB RAS and Novosibirsk State University, Novosibirsk 630090}
\affiliation{Faculty of Mathematics and Physics, Charles University, 121 16 Prague}
%%%\affiliation{Chiba University, Chiba 263-8522}
\affiliation{University of Cincinnati, Cincinnati, Ohio 45221}
\affiliation{Deutsches Elektronen--Synchrotron, 22607 Hamburg}
%%%\affiliation{Department of Physics, Fu Jen Catholic University, Taipei 24205}
\affiliation{Justus-Liebig-Universit\"at Gie\ss{}en, 35392 Gie\ss{}en}
\affiliation{Gifu University, Gifu 501-1193}
%%%\affiliation{II. Physikalisches Institut, Georg-August-Universit\"at G\"ottingen, 37073 G\"ottingen}
%%%\affiliation{The Graduate University for Advanced Studies, Hayama 240-0193}
%%%\affiliation{Gyeongsang National University, Chinju 660-701}
\affiliation{Hanyang University, Seoul 133-791}
\affiliation{University of Hawaii, Honolulu, Hawaii 96822}
\affiliation{High Energy Accelerator Research Organization (KEK), Tsukuba 305-0801}
\affiliation{Hiroshima Institute of Technology, Hiroshima 731-5193}
\affiliation{IKERBASQUE, Basque Foundation for Science, 48011 Bilbao}
%%%\affiliation{University of Illinois at Urbana-Champaign, Urbana, Illinois 61801}
\affiliation{Indian Institute of Technology Guwahati, Assam 781039}
\affiliation{Indian Institute of Technology Madras, Chennai 600036}
%%%\affiliation{Indiana University, Bloomington, Indiana 47408}
\affiliation{Institute of High Energy Physics, Chinese Academy of Sciences, Beijing 100049}
\affiliation{Institute of High Energy Physics, Vienna 1050}
\affiliation{Institute for High Energy Physics, Protvino 142281}
%%%\affiliation{Institute of Mathematical Sciences, Chennai 600113}
\affiliation{INFN - Sezione di Torino, 10125 Torino}
\affiliation{Institute for Theoretical and Experimental Physics, Moscow 117218}
\affiliation{J. Stefan Institute, 1000 Ljubljana}
\affiliation{Kanagawa University, Yokohama 221-8686}
\affiliation{Institut f\"ur Experimentelle Kernphysik, Karlsruher Institut f\"ur Technologie, 76131 Karlsruhe}
%%%\affiliation{Kavli Institute for the Physics and Mathematics of the Universe (WPI), University of Tokyo, Kashiwa 277-8583}
\affiliation{Korea Institute of Science and Technology Information, Daejeon 305-806}
\affiliation{Korea University, Seoul 136-713}
\affiliation{Kyoto University, Kyoto 606-8502}
\affiliation{Kyungpook National University, Daegu 702-701}
\affiliation{\'Ecole Polytechnique F\'ed\'erale de Lausanne (EPFL), Lausanne 1015}
\affiliation{Faculty of Mathematics and Physics, University of Ljubljana, 1000 Ljubljana}
\affiliation{Luther College, Decorah, Iowa 52101}
\affiliation{University of Maribor, 2000 Maribor}
\affiliation{Max-Planck-Institut f\"ur Physik, 80805 M\"unchen}
\affiliation{School of Physics, University of Melbourne, Victoria 3010}
\affiliation{Moscow Physical Engineering Institute, Moscow 115409}
\affiliation{Moscow Institute of Physics and Technology, Moscow Region 141700}
\affiliation{Graduate School of Science, Nagoya University, Nagoya 464-8602}
\affiliation{Kobayashi-Maskawa Institute, Nagoya University, Nagoya 464-8602}
%%%\affiliation{Nara University of Education, Nara 630-8528}
\affiliation{Nara Women's University, Nara 630-8506}
\affiliation{National Central University, Chung-li 32054}
\affiliation{National United University, Miao Li 36003}
\affiliation{Department of Physics, National Taiwan University, Taipei 10617}
\affiliation{H. Niewodniczanski Institute of Nuclear Physics, Krakow 31-342}
\affiliation{Nippon Dental University, Niigata 951-8580}
\affiliation{Niigata University, Niigata 950-2181}
\affiliation{University of Nova Gorica, 5000 Nova Gorica}
\affiliation{Osaka City University, Osaka 558-8585}
%%%\affiliation{Osaka University, Osaka 565-0871}
\affiliation{Pacific Northwest National Laboratory, Richland, Washington 99352}
\affiliation{Panjab University, Chandigarh 160014}
\affiliation{Peking University, Beijing 100871}
\affiliation{University of Pittsburgh, Pittsburgh, Pennsylvania 15260}
%%%\affiliation{Punjab Agricultural University, Ludhiana 141004}
\affiliation{Research Center for Electron Photon Science, Tohoku University, Sendai 980-8578}
%%%\affiliation{Research Center for Nuclear Physics, Osaka University, Osaka 567-0047}
%%%\affiliation{RIKEN BNL Research Center, Upton, New York 11973}
%%%\affiliation{Saga University, Saga 840-8502}
\affiliation{University of Science and Technology of China, Hefei 230026}
\affiliation{Seoul National University, Seoul 151-742}
%%%\affiliation{Shinshu University, Nagano 390-8621}
\affiliation{Soongsil University, Seoul 156-743}
\affiliation{Sungkyunkwan University, Suwon 440-746}
\affiliation{School of Physics, University of Sydney, NSW 2006}
\affiliation{Tata Institute of Fundamental Research, Mumbai 400005}
\affiliation{Excellence Cluster Universe, Technische Universit\"at M\"unchen, 85748 Garching}
\affiliation{Toho University, Funabashi 274-8510}
\affiliation{Tohoku Gakuin University, Tagajo 985-8537}
\affiliation{Tohoku University, Sendai 980-8578}
\affiliation{Department of Physics, University of Tokyo, Tokyo 113-0033}
\affiliation{Tokyo Institute of Technology, Tokyo 152-8550}
\affiliation{Tokyo Metropolitan University, Tokyo 192-0397}
\affiliation{Tokyo University of Agriculture and Technology, Tokyo 184-8588}
\affiliation{University of Torino, 10124 Torino}
%%%\affiliation{Toyama National College of Maritime Technology, Toyama 933-0293}
\affiliation{CNP, Virginia Polytechnic Institute and State University, Blacksburg, Virginia 24061}
\affiliation{Wayne State University, Detroit, Michigan 48202}
\affiliation{Yamagata University, Yamagata 990-8560}
\affiliation{Yonsei University, Seoul 120-749}
  \author{Y.~Kato}\affiliation{Graduate School of Science, Nagoya University, Nagoya 464-8602} % Nagoya
  \author{T.~Iijima}\affiliation{Kobayashi-Maskawa Institute, Nagoya University, Nagoya 464-8602}\affiliation{Graduate School of Science, Nagoya University, Nagoya 464-8602} % Nagoya
  \author{I.~Adachi}\affiliation{High Energy Accelerator Research Organization (KEK), Tsukuba 305-0801} % KEK
% \author{K.~Adamczyk}\affiliation{H. Niewodniczanski Institute of Nuclear Physics, Krakow 31-342} % Krakow
  \author{H.~Aihara}\affiliation{Department of Physics, University of Tokyo, Tokyo 113-0033} % Tokyo
% \author{K.~Arinstein}\affiliation{Budker Institute of Nuclear Physics SB RAS and Novosibirsk State University, Novosibirsk 630090} % BINP
% \author{Y.~Arita}\affiliation{Graduate School of Science, Nagoya University, Nagoya 464-8602} % Nagoya
  \author{D.~M.~Asner}\affiliation{Pacific Northwest National Laboratory, Richland, Washington 99352} % PNNL
% \author{T.~Aso}\affiliation{Toyama National College of Maritime Technology, Toyama 933-0293} % Toyama
% \author{V.~Aulchenko}\affiliation{Budker Institute of Nuclear Physics SB RAS and Novosibirsk State University, Novosibirsk 630090} % BINP
  \author{T.~Aushev}\affiliation{Institute for Theoretical and Experimental Physics, Moscow 117218} % ITEP
% \author{T.~Aziz}\affiliation{Tata Institute of Fundamental Research, Mumbai 400005} % Tata
  \author{A.~M.~Bakich}\affiliation{School of Physics, University of Sydney, NSW 2006} % Sydney
  \author{A.~Bala}\affiliation{Panjab University, Chandigarh 160014} % Panjab
  \author{Y.~Ban}\affiliation{Peking University, Beijing 100871} % Peking
% \author{E.~Barberio}\affiliation{School of Physics, University of Melbourne, Victoria 3010} % Melbourne
% \author{M.~Barrett}\affiliation{University of Hawaii, Honolulu, Hawaii 96822} % Hawaii
% \author{W.~Bartel}\affiliation{Deutsches Elektronen--Synchrotron, 22607 Hamburg} % DESY
% \author{A.~Bay}\affiliation{\'Ecole Polytechnique F\'ed\'erale de Lausanne (EPFL), Lausanne 1015} % Lausanne
% \author{I.~Bedny}\affiliation{Budker Institute of Nuclear Physics SB RAS and Novosibirsk State University, Novosibirsk 630090} % BINP
% \author{P.~Behera}\affiliation{Indian Institute of Technology Madras, Chennai 600036} % IITM
% \author{M.~Belhorn}\affiliation{University of Cincinnati, Cincinnati, Ohio 45221} % Cincinnati
% \author{K.~Belous}\affiliation{Institute for High Energy Physics, Protvino 142281} % Protvino
  \author{V.~Bhardwaj}\affiliation{Nara Women's University, Nara 630-8506} % Nara
% ...
% Karlsruhe
% \author{N.~Zwahlen}\affiliation{\'Ecole Polytechnique F\'ed\'erale de Lausanne (EPFL), Lausanne 1015} % Lausanne
% \author{O.~Zyukova}\affiliation{Budker Institute of Nuclear Physics SB RAS and Novosibirsk State University, Novosibirsk 630090} % BINP
\collaboration{The Belle Collaboration}

%% end author list

\begin{abstract}
We report results of a study of doubly charmed baryons and charmed 
strange baryons.
The analysis is performed using a 980 fb$^{-1}$ data sample collected with the
Belle detector at the KEKB asymmetric-energy $e^{+}e^{-}$ collider.
We search for doubly charmed baryons $\Xi_{cc}^{+(+)}$ 
with the $\Lambda_{c}^{+}K^{-}\pi^{+}(\pi^{+})$ and 
$\Xi_{c}^{0}\pi^{+}(\pi^{+})$ final states. No significant signal is observed. 
We also search for two excited charmed strange baryons, 
$\Xi_{c}(3055)^{+}$ and $\Xi_{c}(3123)^{+}$ with 
the $\Sigma_{c}^{++}(2455)K^{-}$ and $\Sigma_{c}^{++}(2520)K^{-}$ final states. 
The $\Xi_{c}(3055)^{+}$ signal is observed with a significance 
of 6.6 standard deviations including systematic uncertainty,
while no signature of the $\Xi_{c}(3123)^{+}$ is seen. We also study 
properties of the $\Xi_{c}(2645)^{+}$ and measure a width of
2.6 $\pm$ 0.2 (stat) $\pm$ 0.4 (syst) MeV/${\it c}$$^{2}$,
which is the first significant determination.

\end{abstract}


\maketitle

%%%% >>>> keep the final version single-spaced
\tighten

{\renewcommand{\thefootnote}{\fnsymbol{footnote}}}
\setcounter{footnote}{0}

\section{Introduction}\label{section_intro}
In recent years, there has been significant progress in  
charmed baryon spectroscopy, mainly by the Belle and BaBar 
experiments~\citep{Mizuk:2004yu,Chistov:2006zj,Aubert:2006je,Aubert:2006sp,Abe:2006rz,Aubert:2007dt,Solovieva:2008fw,Lesiak:2008wz}.
In particular, all the ground states of the single-charmed baryons predicted by the 
constituent quark model and several excited states 
have been observed~\cite{PDG}. 

However, there are no experimentally established doubly-charmed baryons. 
The lightest doubly-charmed baryon contains two charm quarks
and one up or down quark ($\Xi_{cc}^{+}=\mathnormal{ccd}$, $\Xi_{cc}^{++}=\mathnormal{ccu}$), 
and the spin-parity of the ground state is expected to be $\frac{1}{2}^{+}$.
The mass of the $\Xi_{cc}$ has been extensively studied theoretically, and the 
prediction of the quark model ranges from
3.48 GeV/${\it c}$$^{2}$ to 3.74 GeV/${\it c}$$^{2}$
\citep{Roncaglia:1995az,Ebert:1996ec,SilvestreBrac:1996wp,Tong:1999qs,Gerasyuta:1999pc,Itoh:2000um,Kiselev:2001fw,Narodetskii:2002ib,Ebert:2002ig,Vijande:2004at,Migura:2006ep,Albertus:2006ya,Roberts:2007ni}, whereas the mass 
predicted by lattice QCD ranges from 3.51 GeV/${\it c}$$^{2}$ to 
3.67 GeV/${\it c}$$^{2}$ \citep{Lewis:2001iz,Na:2008hz,Liu:2009jc,Namekawa:2012mp,Alexandrou:2012xk}.
The cross sections of the $\Xi_{cc}$ production in the process 
$e^{+}e^{-}\to \Xi_{cc} X$  at $\sqrt{s}=10.58$ GeV,
where $X$ denotes the remaining particles produced in the fragmentation, 
is predicted to be 70 fb in Ref.~\citep{Kiselev:1994pu} and 230 fb in Ref.~\citep{Ma:2003zk}. 
The cross section of the pair production of the $cc$ and $\bar{c}\bar{c}$ 
diquarks is predicted to be 7 fb~\cite{Braguta:2002qu}.

There have been several experimental studies to search for the $\Xi_{cc}$.
The SELEX collaboration reported evidence for the $\Xi_{cc}^{+}$ in the 
$\Lambda_{c}^{+}K^{-}\pi^{+}$~\cite{Mattson:2002uw} and 
$pD^{+}K^{-}$~\cite{Ocherashvili:2004hi} final states with a mass of about 
3.52 GeV/${\it c}$$^{2}$
using a 600 GeV/${\it c}$ charged hyperon beam.
However, the results have not been supported by FOCUS~\cite{Ratti:2003ez}, 
BaBar~\cite{Aubert:2006qw}, Belle~\cite{Chistov:2006zj} nor LHCb~\cite{Aaij:2013voa}.
The BaBar collaboration searched for the $\Xi_{cc}^{+(+)}$ in the 
$\Lambda_{c}^{+}K^{-}\pi^{+}(\pi^{+})$ and 
$\Xi_{c}^{0}\pi^{+}(\pi^{+})$ decay modes with a 232 fb$^{-1}$ data sample of $e^{+}e^{-}$ collisions at or near the
$\Upsilon (4S)$. 
They found no evidence for
the $\Xi_{cc}^{+(+)}$ and set an upper limit on the product of the 
production cross section and branching fractions of $\Xi_{cc}$ and 
$\Lambda_{c}^{+}$ or $\Xi_{c}^{0}$
to be a few fb, depending on the decay mode.
In our search for the $\Xi_{cc}^{+}$ in the 
$\Lambda_{c}^{+}K^{-}\pi^{+}$ final state with a 462 fb$^{-1}$ data sample of Belle 
at or near the $\Upsilon(4S)$~\cite{Chistov:2006zj}, Belle also found no evidence for the $\Xi_{cc}^{+}$
and set an upper limit on $\sigma(e^{+}e^{-} \to \Xi_{cc}^{+} X)$
$\times$ $\cal B$($\Xi_{cc}^{+} \to \Lambda_{c}^{+}K^{-}\pi^{+}$)/$\sigma(e^{+}e^{-} \to \Lambda_{c}^{+}X)$
of 1.5 $\times$ 10$^{-4}$ with a $p^{\ast}(\Lambda_{c}^{+})>$ 2.5 GeV/${\it c}$ requirement.
Here, $p^{\ast}(\Lambda_{c}^{+})$ is the momentum of the $\Lambda_{c}^{+}$ 
in the center-of-mass (CM) frame.

In this paper, we report on an improved search for the $\Xi_{cc}$ in its weak decays 
to the $\Lambda_{c}^{+}K^{-}\pi^{+}(\pi^{+})$ and $\Xi_{c}^{0}\pi^{+}(\pi^{+})$ 
final states.
The Belle collaboration has collected a data sample with a total 
integrated luminosity of 980 fb$^{-1}$,
which is around two (four) times the statistics of the previous $\Xi_{cc}$ search 
by Belle~\cite{Chistov:2006zj} (BaBar~\cite{Aubert:2006qw}) and supersedes the results in Ref.~\cite{Chistov:2006zj}.
Furthermore, in the previous studies, the $\Lambda_{c}^{+}$ and the 
$\Xi_{c}^{0}$ states have been reconstructed only from 
decay modes of $pK^{-}\pi^{+}$ and $\Xi^{-}\pi^{+}$, respectively. 
We incorporate additional decay modes 
to improve the statistical sensitivity.

%It is possible to increase the statistics
%by including other decay modes as shown in following sections.

The same data sample can be used to study charmed strange baryons, as the 
$\Lambda_{c}^{+}K^{-}\pi^{+}$ and the $\Xi_{c}^{0}\pi^{+}$ 
final states are strong decay modes of excited 
$\Xi_{c}^{+}$ ($\Xi_{c}^{\ast +}$) states.
The BaBar collaboration found two $\Xi_c^{\ast +}$ states, 
$\Xi_c(3055)^{+}$ and $\Xi_c(3123)^{+}$, 
decaying to the $\Lambda_{c}^{+}K^{-}\pi^{+}$ final state through intermediate 
$\Sigma_c(2455)^{++}$ or $\Sigma_c(2520)^{++}$ states
using a data sample of 384 fb$^{-1}$~\cite{Aubert:2007dt}.
Their statistical significance was 6.4 standard deviations ($\sigma$) and 
3.6$\sigma$, respectively. 
A confirmation of these states in other experiments is necessary. 
The $\Xi_{c}^{0}\pi^{+}$
is a strong decay mode of the $\Xi_{c}(2645)^{+}$.
Currently, only the upper limit of 3.1 MeV/${\it c}$$^{2}$ exists for its 
width~\cite{Gibbons:1996yv}.
In this paper, we also report on a search for the $\Xi_c(3055)^{+}$ and 
$\Xi_c(3123)^{+}$ in the
$\Lambda_c^{+}K^{-}\pi^{+}$ final state, and the measurement of the width 
of the $\Xi_{c}(2645)^{+}$.

The remaining sections of the paper are organized as follows.
In section \ref{section_data}, the data samples and the Belle detector 
are described. In section \ref{section_lambdac}, a study of the 
final states with $\Lambda_c^{+}$, {\textit i.e.}, the $\Xi_{cc}$
search and the $\Xi_{c}(3055)^{+}$ and $\Xi_{c}(3123)^{+}$ search, are reported.
In section \ref{section_xic}, a study of the final state with 
$\Xi_c^{0}$ {\textit i.e.}, the $\Xi_{cc}$ search
and measurement of the width of the $\Xi_c(2645)^{+}$, are described.
Finally, conclusions are given 
in section \ref{section_conclusion}.

\section{Data samples and the Belle detector}\label{section_data}
We use a data sample with a total integrated luminosity of 980 fb$^{-1}$ 
recorded with the Belle detector at the KEKB asymmetric-energy 
$e^{+}e^{-}$ collider~\cite{KEKB}. The data samples with 
different beam energies at or near the $\Upsilon(1S)$ to $\Upsilon(5S)$ are combined in this study. The 
beam energies and integrated luminosities are summarized 
in Table \ref{summary_luminosity}. The luminosity-weighted average of $\sqrt{s}$ is 10.59 GeV.

The Belle detector is a large-solid-angle magnetic
spectrometer that consists of a silicon vertex detector (SVD),
a 50-layer central drift chamber (CDC), an array of
aerogel threshold Cherenkov counters (ACC),  % <- \v{C}erenkov 2007.08
a barrel-like arrangement of time-of-flight
scintillation counters (TOF), and an electromagnetic calorimeter
comprised of CsI(Tl) crystals (ECL) located inside 
a superconducting solenoid coil that provides a 1.5~T
magnetic field.  An iron flux return located outside of
the coil is instrumented to detect $K_L^0$ mesons and to identify
muons (KLM).  The detector is described in detail elsewhere~\cite{Belle}.
Two inner detector configurations were used. A 2.0 cm radius beampipe
and a 3-layer silicon vertex detector were used for the first sample
of 156 fb$^{-1}$, while a 1.5 cm radius beampipe, a 4-layer
silicon detector and a small-cell inner drift chamber were used to record
the remaining 824 fb$^{-1}$~\cite{svd2}.\par
The selection of charged hadrons is based on information from the 
tracking system (SVD and CDC) and 
hadron identification system (CDC, ACC, and TOF). The charged proton, 
kaon, and pion that is not 
associated with long-lived particles like $K_{S}^{0}$, 
$\Lambda$ and $\Xi^{-}$, is required to
have a point of closest approach to the interaction point
that is within 0.2 cm in the transverse ($r$-$\phi$) direction and 
within 2 cm along the $z$-axis. (The $z$-axis is opposite
the positron beam direction.) For each track, the likelihood values 
$\mathcal{L}_{p}$, $\mathcal{L}_{K}$, and $\mathcal{L}_{\pi}$ are provided 
for the assumption of proton, kaon and pion, respectively, 
from the hadron identification system, based on 
the ionization energy loss in the CDC, the number of detected Cherenkov 
photons in the ACC, and the time of flight measured by the the TOF.
The likelihood ratio is defined as 
$\mathcal{L}(i:j)=\mathcal{L}_{i}/(\mathcal{L}_{i}+\mathcal{L}_{j})$ and
a track is identified as a proton if the likelihood ratios 
$\mathcal{L}(p:\pi)$ and $\mathcal{L}(p:K)$ are greater than 0.6.
A track is identified as a kaon if the likelihood ratios 
$\mathcal{L}(K:\pi)$ and $\mathcal{L}(K:p)$
are greater than 0.6.
A track is identified as a pion if the likelihood ratios 
$\mathcal{L}(\pi:K)$ and $\mathcal{L}(\pi:p)$
are greater than 0.6.
In addition, electron ($\mathcal{L}_{e}$)  likelihood
is provided based on information from the ECL, ACC, and CDC~\cite{eid}.
A track with an electron likelihood greater than 0.95 is rejected. \par
The momentum averaged efficiencies of hadron identification are about 90$\%$, 90$\%$, and 93$\%$ for pions,
kaons and protons, respectively.
The momentum averaged probability to misidentify a pion (kaon) track as a kaon (pion) track is about 9 (10)$\%$, and the momentum averaged probability to
misidentify a pion or kaon track as a proton track is about 5$\%$.

We use a Monte-Carlo (MC) simulation events generated with EVTGEN~\cite{evtgen}, JETSET~\cite{jetset} 
with final QED final state radiation by PHOTOS~\cite{photos} and then processed by a 
GEANT3~\cite{GEANT} based detector simulation to obtain the reconstruction efficiency and the mass resolution.
\begin{center}
  \begin{table*}[htbp]
    \caption{Summary of the integrated luminosities and beam energies.}
    \begin{tabular}{l|lllll} \hline \hline 
     $\sqrt{s}$                 & $\Upsilon(5S)$/near it &$\Upsilon(4S)$/near it & $\Upsilon(3S)$/near it& $\Upsilon(2S)$/near it & $\Upsilon(1S)$/near it  \\  \hline
     Integrated luminosity (fb$^{-1}$) &121.0/29.3              &702.6/89.5             &2.9/0.3                & 24.9/1.8               & 5.7/1.8    \\  \hline \hline
    \end{tabular}
    \label{summary_luminosity}
  \end{table*}
\end{center}

\section{Final state with the $\Lambda_{c}^{+}$}\label{section_lambdac}
In this section, the analysis using the final states with the $\Lambda_{c}^{+}$ baryon
is described.
Reconstruction of the $\Lambda_{c}^{+}$ candidate is explained first,
followed by the description of the $\Xi_{cc}^{+(+)}$ search in its decay 
into $\Lambda_{c}^{+}K^{-}\pi^{+}(\pi^{+})$ and
the study of two charmed strange baryons, $\Xi_{c}(3055)^{+}$ and $\Xi_{c}(3123)^{+}$.
Throughout this paper, the selection criteria are determined to maximize
the figure of merit (FOM), defined as $\epsilon/\sqrt{N_{\rm bg}}$, where $\epsilon$ is the $\Xi_{cc}$ efficiency 
for the selection criteria and $N_{\rm bg}$ is the number of background events under the signal peak
except for the scaled momentum selection for $\Xi_{c}(3055)^{+}$ and $\Xi_{c}(3123)^{+}$ search, which followed
BaBar's analysis. The distribution of background events is estimated based on data. 
When the selection criteria are determined, we hide the possible signal peak by smearing 
invariant mass of the $\Xi_{cc}$ candidates event by event with a Gaussian having a 50 MeV/${\it c}$$^{2}$ width 
in order to avoid any biases.

\subsection{Reconstruction of the $\Lambda_{c}^{+}$}\label{section_evesele}
The $\Lambda_{c}^{+}$ candidates are reconstructed in the 
$pK^{-}\pi^{+}$ and $pK^{0}_{S}$ decay modes~\cite{CC}.
The $K^{0}_{S}$ candidate is reconstructed from its decay into $\pi^{+}\pi^{-}$.
A pair of oppositely charged pions that have an invariant mass
within 8 MeV/${\it c}$$^{2}$ of the nominal $K^{0}_{S}$ mass, 
which corresponds to approximately 3.5$\sigma$ of the mass resolution, is used.
Two pion tracks are fitted to a common vertex. The result of the fit is used
to suppress misreconstructed $K^{0}_{S}$ candidates and to perform further vertex fit of the $\Lambda_{c}^{+} \to pK^{0}_{S}$.
The vertex of the two pions for the $K^{0}_{S}$ is required to be displaced from the interaction point (IP)
in the direction of the pion pair momentum~\cite{cite:ks}.
The daughters of the $\Lambda_{c}^{+}$ are fitted to a common vertex;
the invariant mass of the daughters must be within 5 (6) MeV/${\it c}$$^{2}$, or 1.5$\sigma$,
nominal $\Lambda_{c}^{+}$ mass for the $pK^{-}\pi^{+}$ ($pK^0_{S}$) decay mode.
The $\chi^{2}$ value of the common vertex fit of the $\Lambda_{c}^{+}$ is required to be less than 50.
For the remaining candidate, a mass constraint fit to the $\Lambda_{c}^{+}$ mass is performed to improve the momentum resolution.
%For the case of $pK^{0}_{S}$ decay mode, the result of common vertex fit of the $K^{0}_{S}$
%is used to extrapolate the $K^{0}_{S}$ track.
As the signal-to-background ratio for the $\Lambda_{c}^{+}$ candidates
is similar for the $pK^{-}\pi^{+}$ and $pK^0_{S}$
decay modes, they are combined in the following analysis. 
By including the $pK^0_{S}$ mode in addition to the $pK^{-}\pi^{+}$ mode,
the yield of the $\Lambda_c^{+}$ is increased by about 20$\%$.

\subsection{Search for doubly charmed baryons in $\Lambda_{c}^{+}K^{-}\pi^{+}(\pi^{+})$} \label{section_xicc_lambdac}
We search for the $\Xi_{cc}^{+(+)}$ in its decay into 
$\Lambda_{c}^{+}K^{-}\pi^{+}(\pi^{+})$ in the mass range of 
3.2-4.0 GeV/${\it c}$$^{2}$. The expected mass resolution of the 
$\Xi_{cc}$ estimated from MC is 2.0-3.5 MeV/${\it c}$$^{2}$, depending
on the mass of the $\Xi_{cc}$ (degrading with increasing mass).
%In order to hide the true mass distribution of $\Xi_{cc}$ candidates, invariant mass of $\Xi_{cc}$
%candidate is smeared by a random Gaussian function with a width of
%50 MeV/c$^{2}$ in the analysis program for each $\Xi_{cc}$ candidate 
%until all the selection criterias were fixed. \par
In order to reduce the combinatorial background, a selection on the scaled momentum $x_{p}$ = $p^{\ast}$/$\sqrt{s/4-m^{2}}$ is used, where 
$p^{\ast}$ is the CM momentum of a $\Xi_{cc}$ candidate and $s$ is CM energy squared and $m$ is mass of the $\Xi_{cc}$ candidate.
As there is no measurement of the $x_{p}$ spectrum for $\Xi_{cc}$ production,
we assume it to be the same as that of the $\Lambda_{c}^{+}$, 
which has been precisely measured~\cite{Aubert:2006cp}.
The $x_{p}$ spectrum is represented by a smooth polynomial function and is used to generate a MC
sample for the $\Xi_{cc}$ signal.
Decays of the $\Xi_{cc}$ and $\Lambda_{c}^{+}$ are generated according to
the available phase space distribution.
The number of background events as a function of the $x_{p}$ cut is estimated 
based on smeared data.
The FOM as a function of the $x_{p}$ cut is surveyed in the search region.
The optimization procedure yields $0.5<x_{p}<1.0$ regardless of the 
$\Xi_{cc}$ mass. To check the validity of our analysis, we independently 
examine the $x_{p}$ spectrum of the $\Lambda_{c}^{+}$ and confirm that it is 
consistent with that presented in Ref. \cite{Aubert:2006cp}.

\begin{figure*}[htbp]
  \begin{center}
    \includegraphics[scale=0.25]{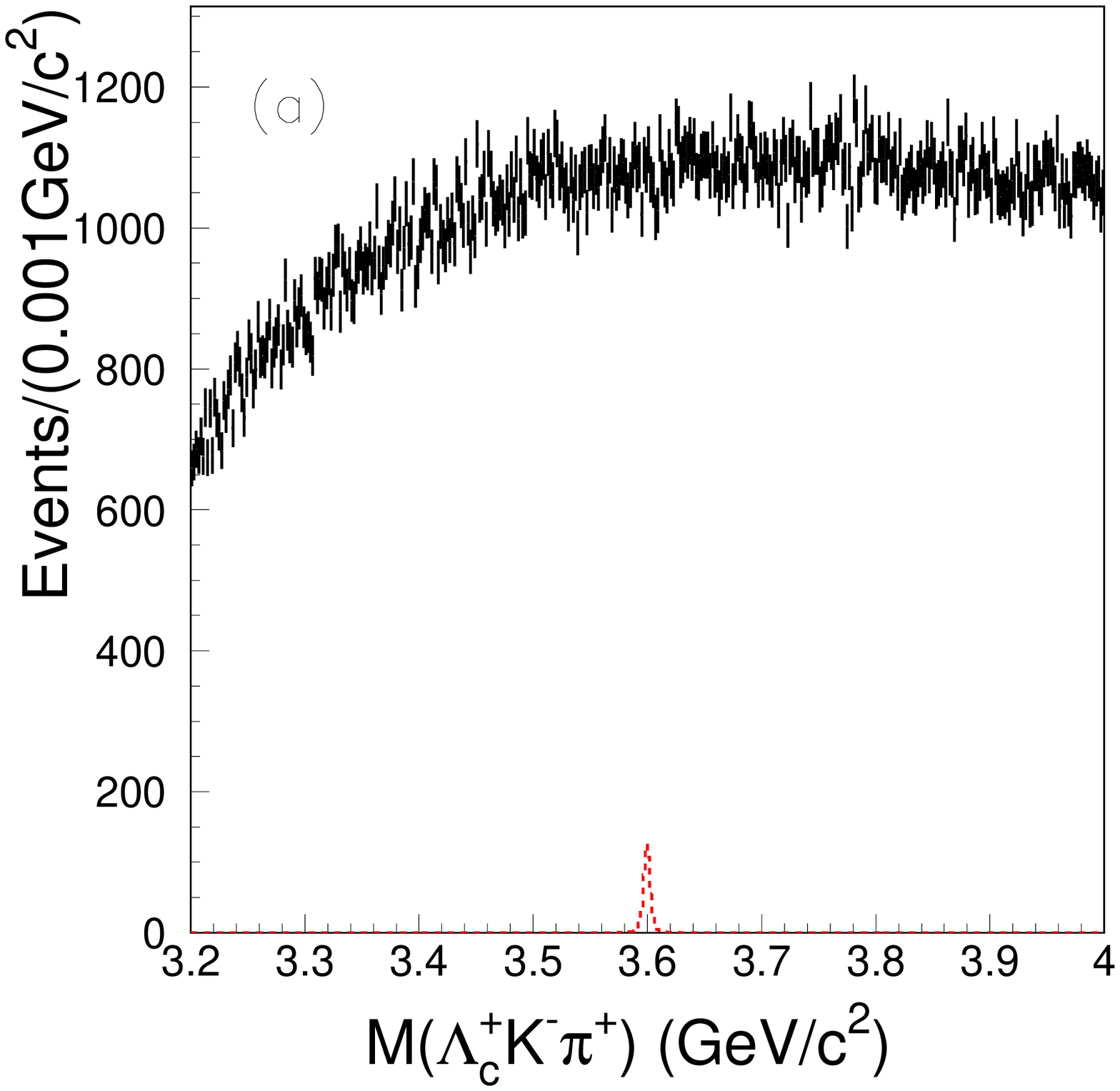}
    \includegraphics[scale=0.25]{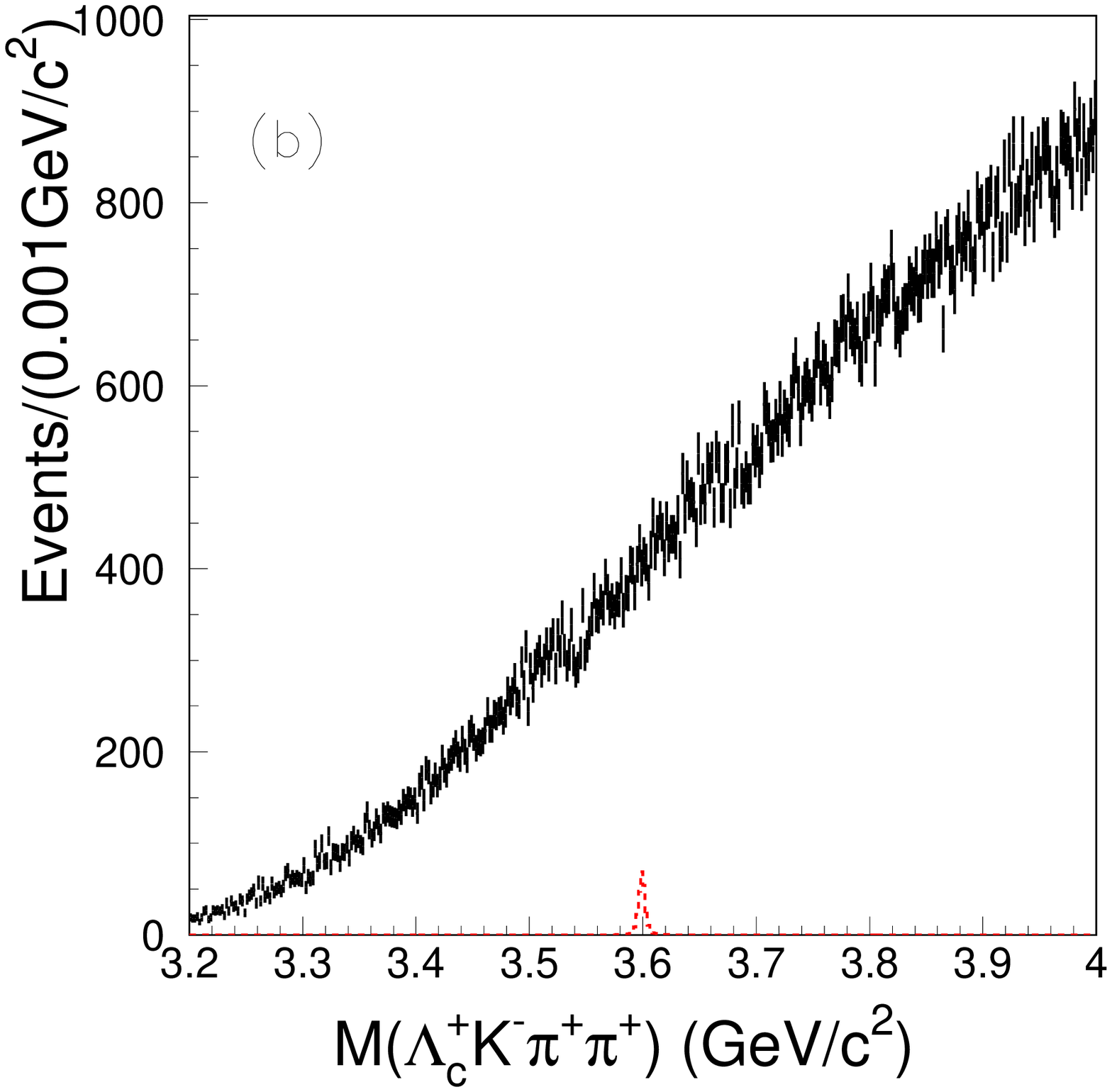}

    \includegraphics[scale=0.25]{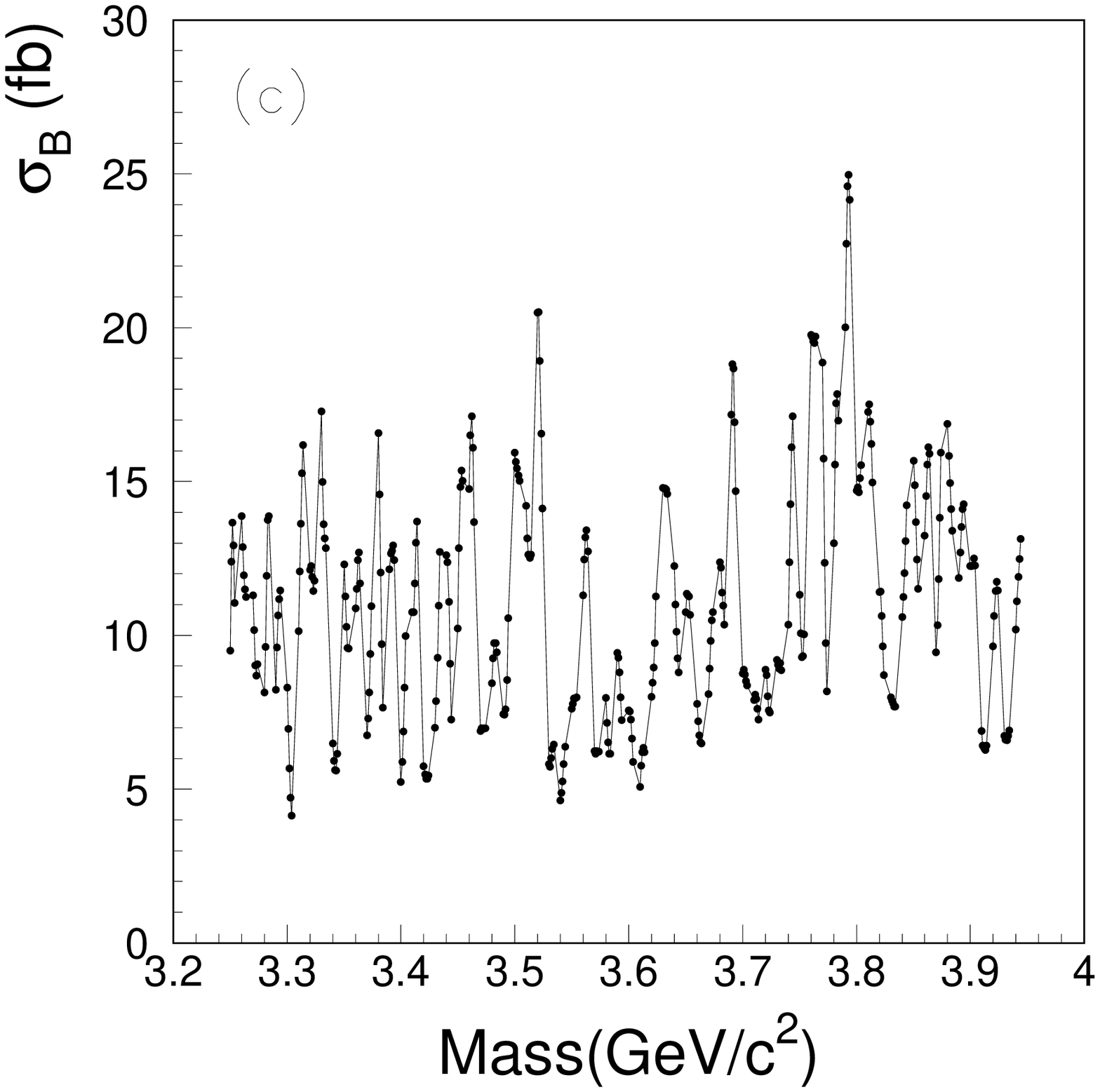}
    \includegraphics[scale=0.25]{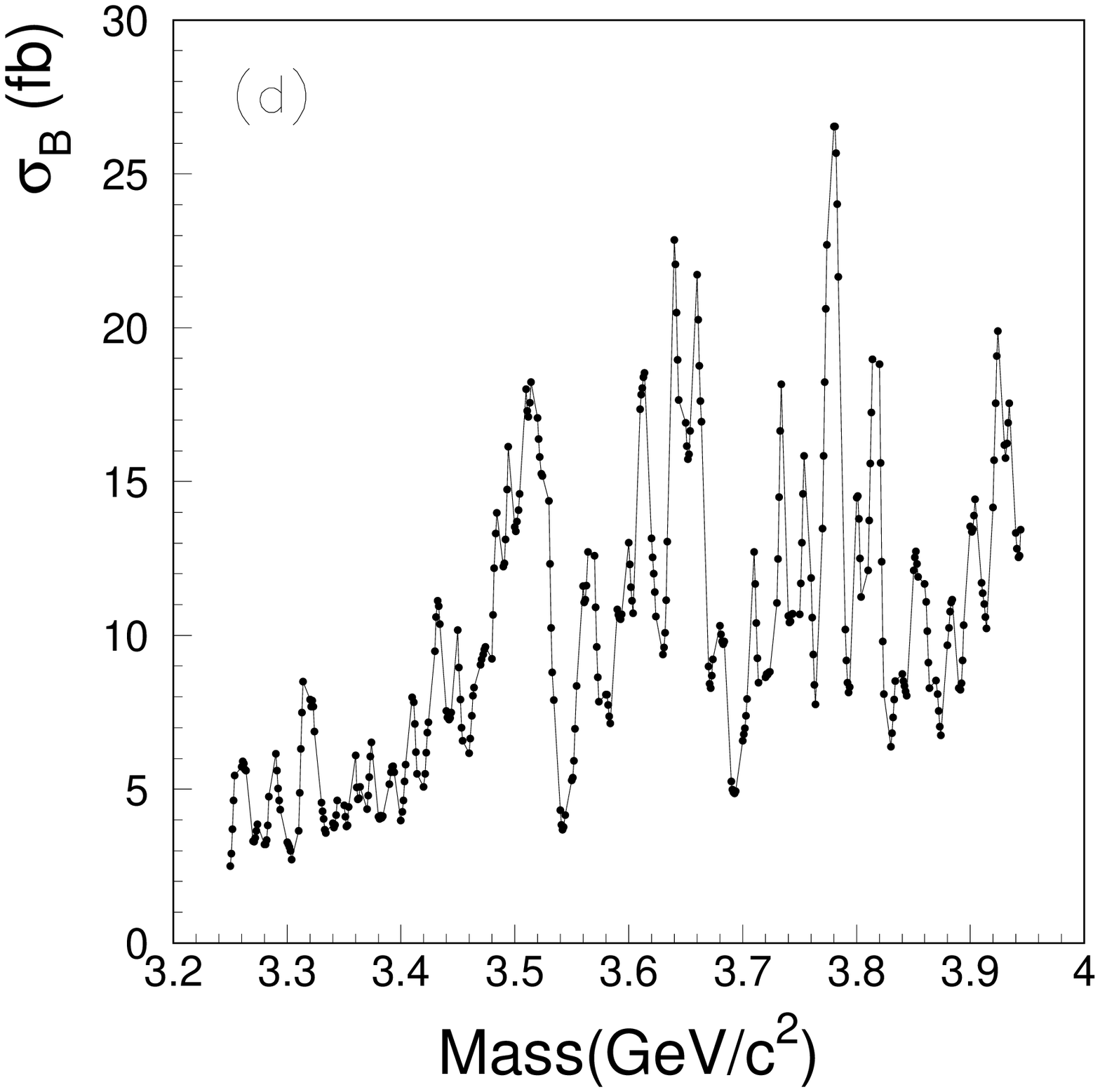}
    \caption{Invariant mass distribution of the $\Xi_{cc}$ candidates for
             (a) $M(\Lambda_{c}^{+}K^{-}\pi^{+})$,
             (b) $M(\Lambda_{c}^{+}K^{-}\pi^{+}\pi^{+})$; the
             vertical error bars are from data and the dashed histogram are from signal MC
             for the $\Xi_{cc}$ signal generated with a mass of 3.60 GeV/${\it c}$$^{2}$ 
             and a production cross section $\sigma(e^{+}e^{-}\to \Xi_{cc}^{+(+)} X)$ of 500 fb and 
             ${\cal B}(\Xi_{cc}^{+(+)} \to \Lambda_{c}^{+}K^{-}\pi^{+}(\pi^{+}))$ of 5$\%$.
            95$\%$ C.L. upper limit of $\sigma_{{\cal B}}$ as a function of the mass with a 1 MeV/${\it c}$$^{2}$ step for 
	    (c) $\Xi_{cc}^{+}$ and (d) $\Xi_{cc}^{++}$.}
    \label{open_lambdac}
  \end{center}
\end{figure*}

Figure \ref{open_lambdac} (a) and (b) show the $M(\Lambda_{c}^{+}K^{-}\pi^{+})$ and 
$M(\Lambda_{c}^{+}K^{-}\pi^{+}\pi^{+})$ distributions, respectively, for data after all the event selections applied.
No significant signal is seen in the data for either 
$\Xi_{cc}^{+}$ or $\Xi_{cc}^{++}$.
The statistical significance for a given mass 
is evaluated with an unbinned extended maximum likelihood (UML) fit.
The probability density function (PDF) for the signal is described with
signal MC generated for each given $\Xi_{cc}$ mass, whereas a third-order 
polynomial function is used as the background PDF.
The statistical significance is defined as 
$\sqrt{-2\ln{(\mathcal{L}_{0}/\mathcal{L})}}$, 
where $\mathcal{L}_{0}$ is the likelihood
for the fit without the signal component and $\mathcal{L}$ is the 
likelihood for the fit with the signal component included.
The significance is evaluated for the $\Xi_{cc}$ mass scanned with a 
1 MeV/${\it c}$$^{2}$ step in the search region.
None of the mass points give a significance exceeding the 3$\sigma$ level.

%The 95$\%$ confidence level (C.L.) upper limit for the
%$\sigma(e^{+}e^{-}\to \Xi_{cc}^{+(+)} X) \times {\cal B}(\Xi_{cc}^{+(+)} \to \Lambda_{c}^{+}K^{-}\pi^{+}(\pi^{+}))$
%produced with the  $0.5<x_{p}<1.0$ condition (denoted as $\sigma_{{\cal B}}$) is
%evaluated. The $\sigma_{\cal B }$ is defined as:
%\begin{equation}\label{equ_sigma}
%\sigma_{\cal B }=\frac{N_{{\rm sig}}}{2L {\cal B} _{pK^{-}\pi^{+}}(\epsilon_{pK^{-}\pi^{+}}+\epsilon_{pK_{S}}{\cal B} _{pK_{S}}/{\cal B}_{pK^{-}\pi^{+}})} ,
%\end{equation}

\begin{figure*}[htbp]
  \begin{center}
    \includegraphics[scale=0.25]{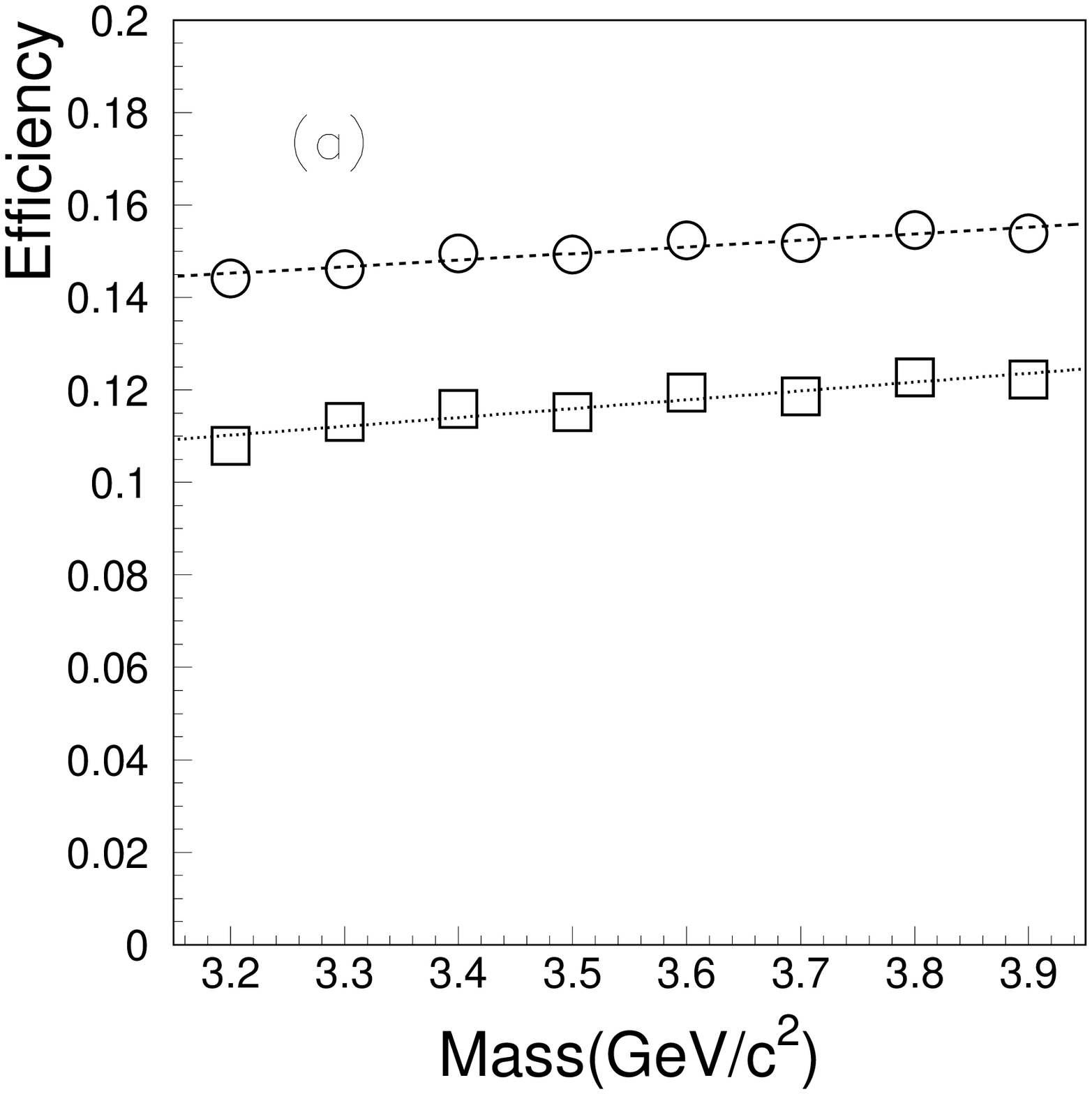}
    \includegraphics[scale=0.25]{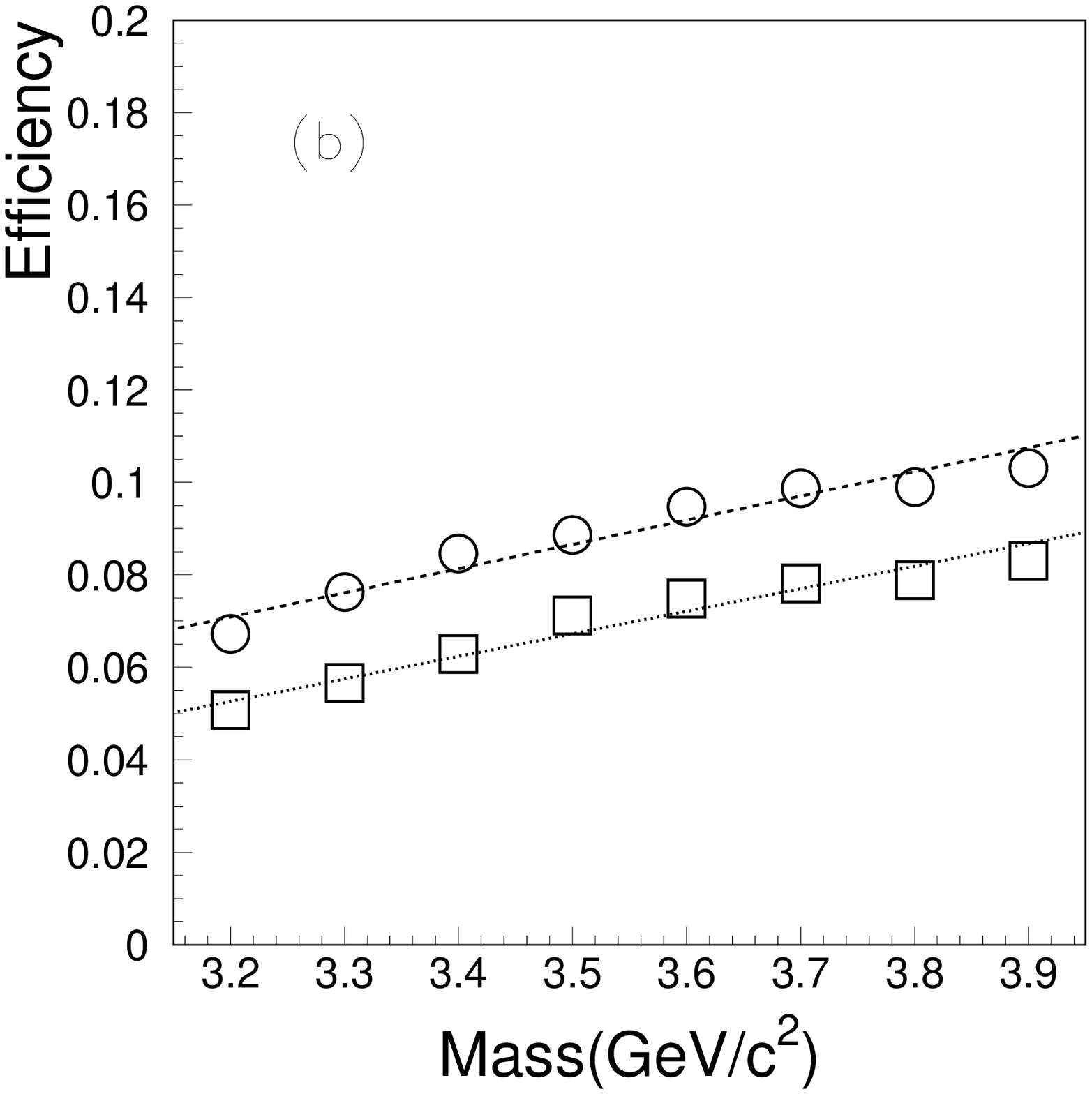}
    \caption{Reconstruction efficiency as a function of the $\Xi_{cc}$ mass, for
            (a) $\Xi_{cc}^{+}$, (b) $\Xi_{cc}^{++}$. Circular points are for $\Lambda_{c}^{+} \to pK^{-}\pi^{+}$ and
             square points are for $\Lambda_{c}^{+} \to pK^{0}_{S}$. The 
lines show the result of the fit with a linear function.}
    \label{eff_lambdackpi}
  \end{center}
\end{figure*}

The 95$\%$ confidence level (C.L.) upper limit for the product of the cross section and branching fraction to the $\Lambda_{c}K^{-}\pi^{+}(\pi^{+})$
state produced with the  $0.5<x_{p}<1.0$ condition,
\begin{eqnarray*}\label{equ_sigma}
\sigma_{\cal B } &\equiv& \sigma(e^{+}e^{-} \to \Xi_{cc}^{+(+)}X) \times {\cal B}(\Xi_{cc}^{+(+)} \to \Lambda_{c}^{+}K^{-}\pi^{+}(\pi^{+})) \\
&=&\frac{N_{{\rm sig}}}{2L {\cal B} _{pK^{-}\pi^{+}}(\epsilon_{pK^{-}\pi^{+}}+\epsilon_{pK_{S}}{\cal B} _{pK_{S}}/{\cal B}_{pK^{-}\pi^{+}})} ,
\end{eqnarray*}
is evaluated. Here, $L$ is the total integrated luminosity, $N_{{\rm sig}}$ is the $\Xi_{cc}$ 
signal yield, 
${\cal B} _{pK^{-}\pi^{+}}$ is the branching fraction of the $\Lambda_{c}^{+}\to pK^{-}\pi^{+}$ (which amounts to 0.050 $\pm$ 0.013),
${\cal B} _{pK_{S}}$ is the branching fraction of $\Lambda_{c}^{+}\to pK^{0}_{S}$ 
measured relative to the $pK^{-}\pi^{+}$ mode 
(${\cal B} _{pK_{S}}$/${\cal B} _{pK^{-}\pi^{+}}$ = 0.24 $\pm$ 0.02), and
$\epsilon_{pK^{-}\pi^{+}(pK_{S})}$ is the reconstruction efficiency for 
the $\Lambda_c^{+}\to pK^{-}\pi^{+}$ ($\Lambda_c^{+}\to pK^{0}_{S}$) 
decay mode evaluated as a function of the $\Xi_{cc}$ mass.
The efficiencies for the $\Xi_{cc}^{+(+)}$ as a function of their masses are shown in Fig. \ref{eff_lambdackpi}.
The factor of two in the denominator comes from inclusion of the charge-conjugate mode.
By including this factor, our measurement can be compared with the
theoretical predictions~\citep{Kiselev:1994pu,Ma:2003zk}; 
while to compare with the prediction in Ref.~\citep{Braguta:2002qu}, 
it is necessary to multiply our $\sigma_{\cal B }$
measurement by 2 because they predicted the cross section of the pair 
production of the $cc$ and $\bar{c}\bar{c}$ diquarks.
In BaBar's measurement \citep{Aubert:2006qw}, they do not 
introduce the factor of two
(\textit{i.e.,} they report an upper limit for the sum of the 
$\sigma(e^{+}e^{-}\to \Xi_{cc}^{+(+)} X)$ and its charge-conjugate mode).
Therefore, our measurement should be doubled when comparing with BaBar's result.
We note that the cross section reported here and elsewhere in this paper is a visible cross section
(\textit{i.e.,} a radiative correction is not applied.

%We note the cross section reported here and elsewhere in this paper is not a Born cross section
%{\textit i.e.}, radiative correction is not applied.
%measurement is not a Born cross section but includes the process with
%initial state radiation. 

The upper limit is evaluated following the Bayesian approach.
First, we scan the likelihood profile by determining the 
likelihood values as a function of the 
$\sigma_{{\cal B} }$ ($\mathcal{L}(\sigma_{{\cal B} })$),
varying $N_{\rm sig}$ from zero up to the $N_{{\rm sig}}$ value for which 
the likelihood drops to zero.
Then, $\mathcal{L}(\sigma_{{\cal B} })$ is convolved with a Gaussian 
whose width equals the systematic uncertainties of $\sigma_{{\cal B}}$.
The $\sigma_{{\cal B}}$ value for which the integral (starting from 
$\sigma_{{\cal B} }=0$) becomes 95$\%$ of the entire area
is regarded as the 95$\%$ C.L. upper limit.

We consider the following systematic uncertainties in the $\Xi_{cc}$ search.
The systematic uncertainty due to the efficiency of pion and kaon identification 
is estimated from the ratio of the yield of the
$D^{\ast +} \to D^{0}\pi^{+}$, $D^{0}\to K^{-}\pi^{+}$ with and without 
the pion/kaon identification requirements for data and MC.
The difference of the ratio between data and MC is corrected and 
the statistical error of the ratio is regarded as the systematic uncertainty.
The systematic uncertainty due to the efficiency of proton identification 
is estimated using the ratio of the yield of the 
$\Lambda \to p \pi^{-}$ with and without the proton identification requirement.
The difference of the ratio between data and MC is corrected and 
the statistical error of the ratio is regarded as the systematic uncertainty.
The systematic uncertainty due to the charged track reconstruction 
efficiency is estimated
using the decay chain $D^{\ast +}\to \pi^{+}D^{0}$, $D^{0}\to \pi^{+}\pi^{-}K^{0}_{S}$, and $K^0_{S}\to\pi^{+}\pi^{-}$,
where $K^{0}_{S} \to \pi^{+} \pi^{-}$ is either partially or fully 
reconstructed. The ratio of the yields
for partially and fully reconstructed signals in data and MC is compared, 
and the difference is taken as the systematic uncertainty. 
This amounts to 0.35$\%$ per track.
The systematic  uncertainty of the total integrated luminosity is 1.4$\%$.
%The systematic error due to the signal PDF is studied from 
%the mass resolution of $\Lambda_{c}^{+}$.   The mass resolution of the $\Lambda_c^{+}$ for data is 5 $\%$ broader than that of the MC.
%The systematic uncertainty is evaluated by an extracting the yield of the signal MC on pseudo background sample
%similar to data with signal PDF and 5 $\%$ narrower PDF in several mass region. The largest difference of 3 $\%$, between two ways, 
%is regarded as systematic error.
To check the systematic error due to the signal PDF, we compare the 
mass resolution of the $\Lambda_c^{+}$ in data and MC.
We find that the resolution for data is 5$\%$ larger than in MC. 
To monitor the effect of this discrepancy, we perform a pseudo-experiment 
test in which we extract the signal yield with 
correct PDF and one that is narrower by 5$\%$. The largest difference of 
3$\%$ measured in this test
is regarded as the systematic  uncertainty.  
The systematic  uncertainty related to the $\Lambda_c^{+}$ branching fraction 
is propagated from the errors taken from the PDG~\cite{PDG}.
To estimate the systematic uncertainty of the reconstruction efficiency due to the possible
difference of $x_{p}$ spectrum between our assumption (the same as that of $\Lambda_{c}^{+}$)
and actual one, we examine the $x_{p}$ dependence of the reconstruction efficiency.
The root mean square of the reconstruction efficiency in the region of 
$0.5<x_{p}<1.0$ is regarded as the systematic uncertainty.  
The elements of the systematic  uncertainty for the measurement of the $\sigma_{{\cal B} }$ 
are enumerated in the first and second columns of Table \ref{summary_sys}.

Figures \ref{open_lambdac} (c) and (d) show the 
95$\%$ C.L. upper limit on $\sigma_{{\cal B}}$ 
for $\Xi_{cc}^{+}$ and $\Xi_{cc}^{++}$, respectively, as a function of the mass with a 
1 MeV/${\it c}$$^{2}$ step.
The upper limit is in the range of 4.1--25.0 fb for the $\Xi_{cc}^{+}$ and 
2.5--26.5 fb for the $\Xi_{cc}^{++}$.

%\begin{center}
%  \begin{table*}[htbp]
%    \caption{Summary of the reconstruction efficiency for the $\Xi_{cc}$.
%             The efficiency becomes higher as the mass becomes higher.}
%    \begin{tabular}{l|l} \hline \hline 
%     mode & efficiency \\  \hline
%     $\Xi_{cc}^{+}$: $\Lambda_{c}^{+} \to pK^{-}\pi^{+}$  & 0.145-0.156 \\
%     $\Xi_{cc}^{+}$: $\Lambda_{c}^{+} \to pK^{0}_{S}$         & 0.110-0.125  \\ 
%     $\Xi_{cc}^{++}$: $\Lambda_{c}^{+}\to pK^{-}\pi^{+}$  & 0.0709-0.113 \\
%     $\Xi_{cc}^{++}$: $\Lambda_{c}^{+}\to pK^{0}_{S}$        & 0.0527-0.0917 \\ \hline \hline
%    \end{tabular}
%    \label{summary_eff_lamkpi}
%  \end{table*}
%\end{center}

\begin{center}
  \begin{table*}[htbp]
    \caption{Summary of the systematic uncertainties in the $\sigma_{{\cal B} }$ and $\sigma_{{\cal B}^{2} }$ measurement (\%).}
%    \begin{tabular}{l|l@{\hspace{1cm}}l@{\hspace{1cm}}l@{\hspace{1cm}}l@{\hspace{1cm}}l@{\hspace{1cm}}l@{\hspace{1cm}}l} \hline \hline
     \begin{tabular}{c|c@{\hspace{1cm}}c@{\hspace{1cm}}c@{\hspace{1cm}}c@{\hspace{1cm}}c} \hline \hline
       Source & $\Xi_{cc}^{+}$ with $\Lambda_{c}^{+}$ & $\Xi_{cc}^{++}$ with $\Lambda_{c}^{+}$ & $\Xi_{c}(3123)^{+}$ & $\Xi_{cc}^{+}$ with $\Xi_{c}^{0}$ & $\Xi_{cc}^{++}$ with $\Xi_{c}^{0}$ \\ \hline
       Particle ID               &2.0  &2.4  & 2.0  & 3.5 & 3.8  \\ 
       Tracking                  &1.8  &2.1  & 1.8  & 1.8 & 2.1  \\ 
       Signal PDF                &3.5  &3.5  & 28.0 & 3.5 & 3.5  \\
       Luminosity                &1.4  &1.4  & 1.4  & 1.4 & 1.4  \\ 
       $\cal{B}$                 &26.0 &26.0 & 1.6  & 0.7 & 0.7  \\ 
       $x_{p}$                   &2.1  &2.3  & 1.2  & 6.0 & 5.7  \\ 
       $N^{i}_{\Xi_c(2645)^{+}}$ & -   & -   & -    & 4.3 & 4.3  \\ \hline
       total                     &26.5 &26.6 & 28.2 & 9.2 & 9.2  \\ \hline \hline
    \end{tabular}
    \label{summary_sys}
  \end{table*}
\end{center}

\subsection{Search for the $\Xi_c^{+}(3055)$ and $\Xi_c^{+}(3123)$} \label{section_xicstar}
\begin{figure*}[htbp]
  \begin{center}
    \includegraphics[scale=0.25]{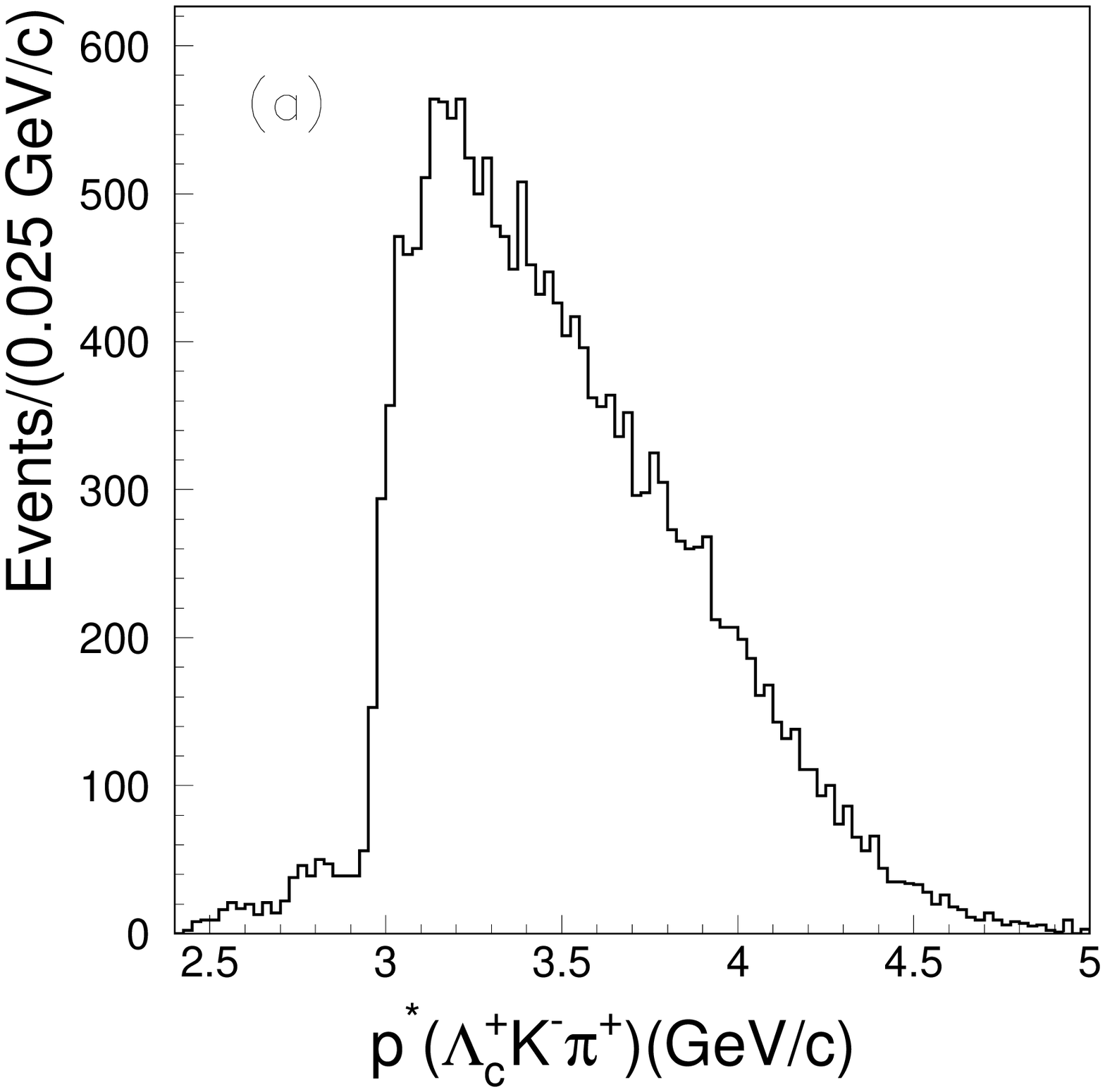}
    \includegraphics[scale=0.25]{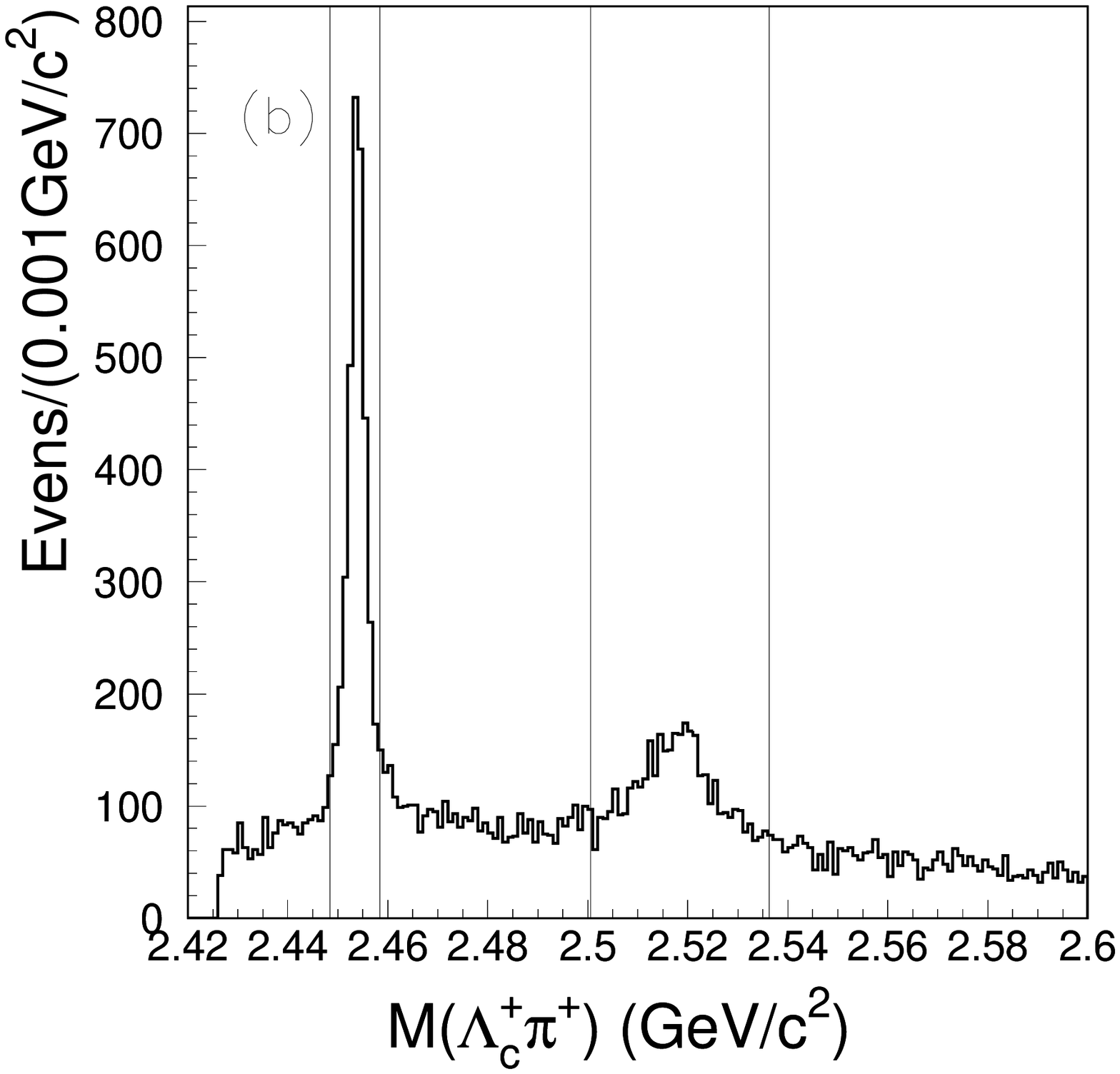}

    \includegraphics[scale=0.25]{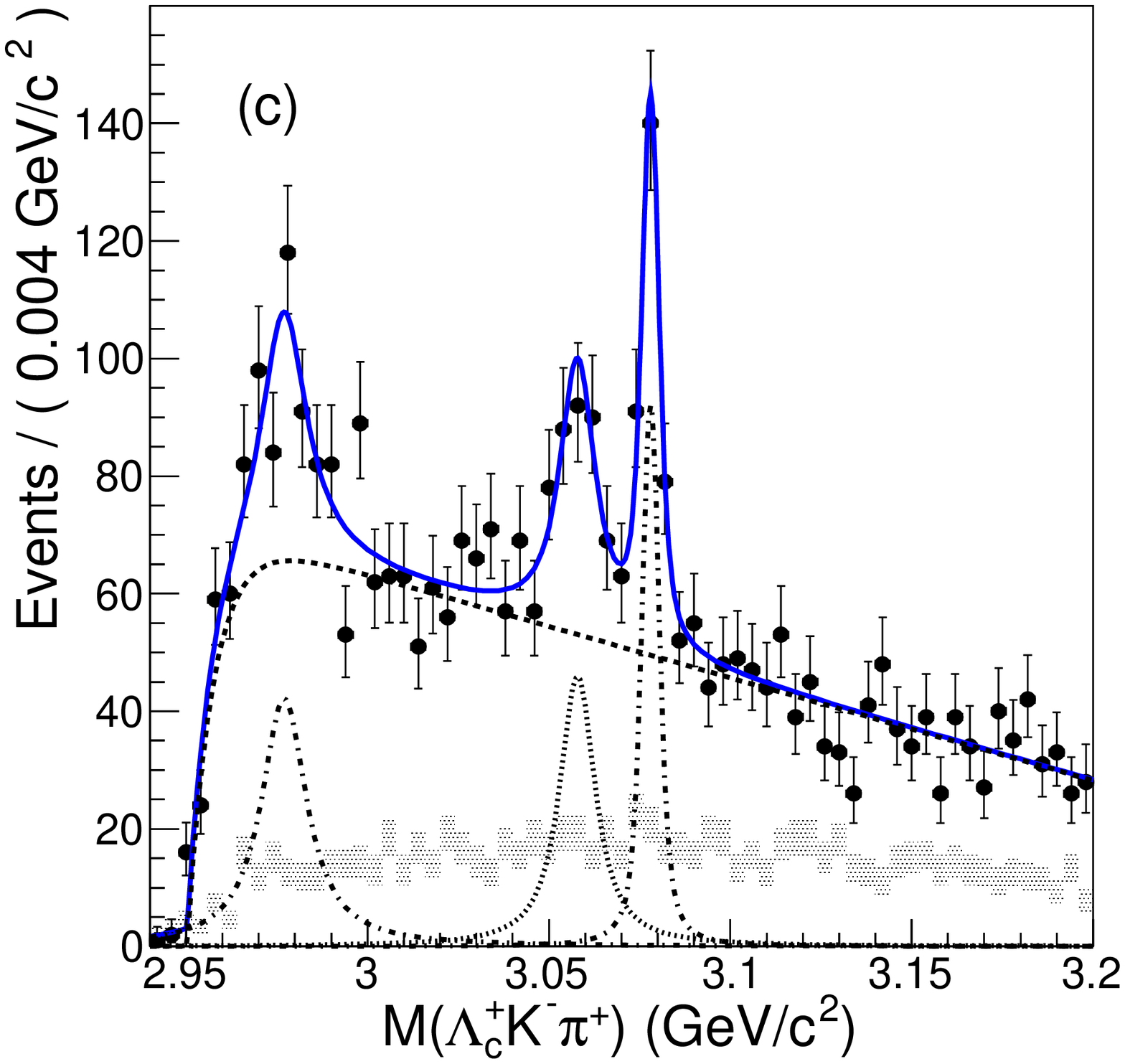}
    \includegraphics[scale=0.25]{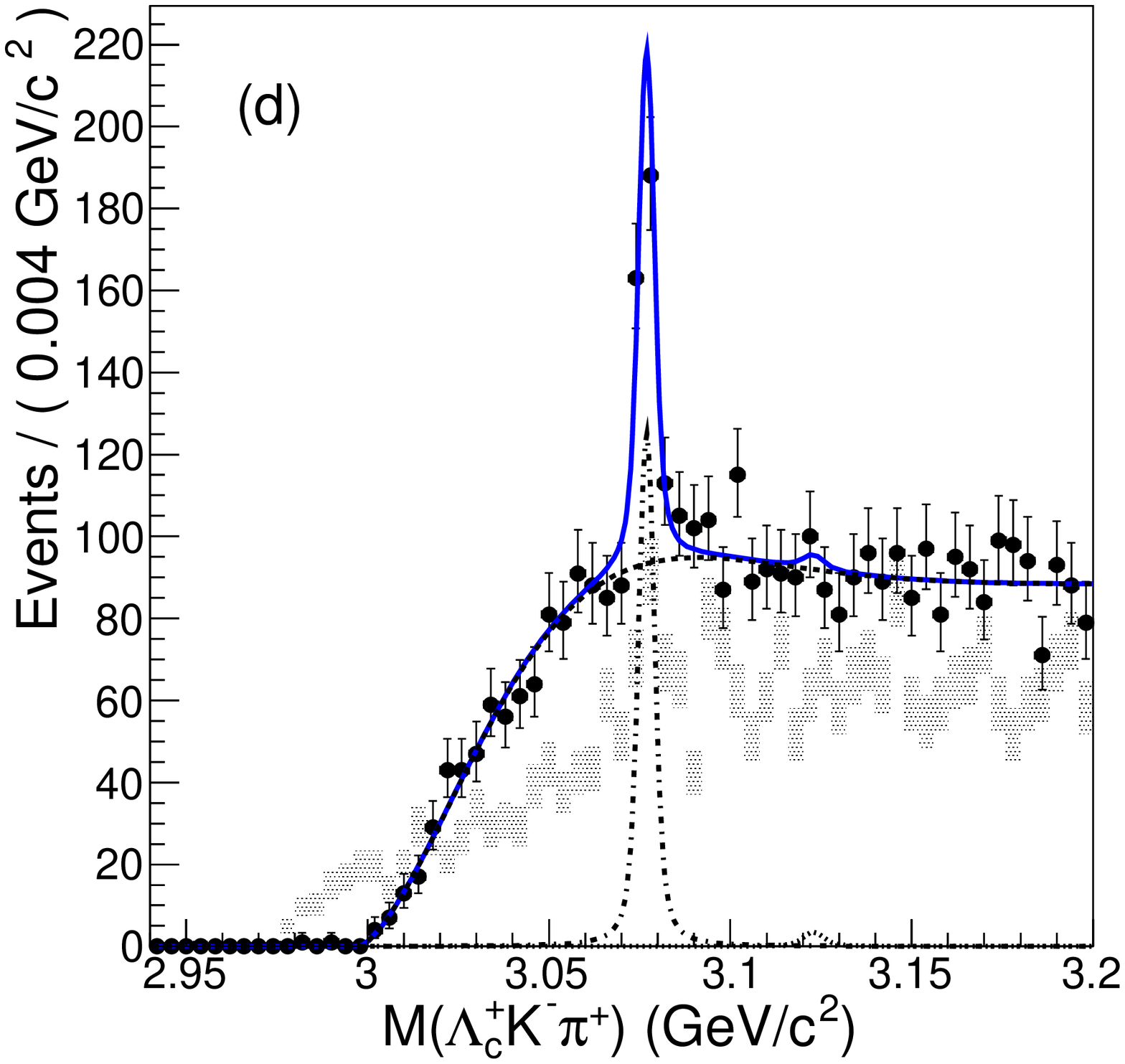}
    \caption{(a) The $p^{\ast}(\Lambda_{c}^{+}K^{-}\pi^{+})$ distribution from data.
             (b) The $M(\Lambda_c^{+}\pi^{+})$ distribution. The vertical 
lines show the selected regions of the $\Sigma_c(2455)^{++}$ and 
                $\Sigma_c(2520)^{++}$.
             (c) The $M(\Lambda_c^{+}K^{-}\pi^{+})$ distribution with 
$\Sigma_c(2455)^{++}$ selection. The dots with error bars show the 
                 distribution for the $\Sigma_c(2455)^{++}$ selected region 
                 whereas the rectangles show the distribution for 
the $\Sigma_c(2455)^{++}$ sideband region.
		 The solid line shows the 
                 fit result. The dashed, dotted, and dash-dotted lines 
show the contributions from 
                 the background, $\Xi_{c}(3055)^{+}$, and $\Xi_{c}(2980)^{+}$ or $\Xi_{c}(3080)^{+}$, respectively.
             (d) The $M(\Lambda_c^{+}K^{-}\pi^{+})$ distribution with $\Sigma_c(2520)^{++}$ selection.
		 The dots with error bars show the 
                 distribution for $\Sigma_c(2520)^{++}$ selected region 
                 whereas the rectangles show the distribution for 
the $\Sigma_c(2520)^{++}$ sideband region.
		 The solid line shows the fit result. The dashed, 
dotted, and dash-dotted lines show the contributions from
              the   background, $\Xi_{c}(3123)^{+}$, and $\Xi_{c}(3080)^{+}$, 
respectively.}
    \label{momdis_lamkpi_xic}
  \end{center}
\end{figure*}

In this section, a search for the $\Xi_c^{+}(3055)$ and 
$\Xi_c^{+}(3123)$ is described.
Here, we require $x_{p}$ to be greater than 0.7. 
In the analysis by BaBar \cite{Aubert:2007dt}, they required 
$p^{\ast}(\Lambda_{c}^{+}K^{-}\pi^{+})>$2.9 GeV/${\it c}$, which is similar
to our $x_{p}$ cut as illustrated by the 
$p^{\ast}(\Lambda_{c}^{+}K^{-}\pi^{+})$ distribution, with
the $x_{p}$  cut and $2.9$ GeV/${\it c}$$^{2}<$ $M(\Lambda_c^{+}K^{-}\pi^{+}) \: <$ 3.2 GeV/${\it c}$$^{2}$ required
as shown in Fig. \ref{momdis_lamkpi_xic} (a). 
Figure \ref{momdis_lamkpi_xic} (b) shows the $M(\Lambda_{c}^{+}\pi^{+})$ 
distribution, where contributions from the
$\Sigma_{c}(2455)^{++}$ and the $\Sigma_{c}(2520)^{++}$ baryons are clearly visible.
We select the $\Sigma_c(2455)^{++}$ ($\Sigma_c(2520)^{++}$) region by 
requiring 
$|M(\Lambda_c^{+}\pi^{+})-m_{\Sigma_c^{++}}| \: <$ 5 (18) MeV/${\it c}$$^{2}$, 
where $m_{\Sigma_c^{++}}$ is the nominal 
mass of the  $\Sigma_c(2455)^{++}$ or $\Sigma_c(2520)^{++}$.

Figure \ref{momdis_lamkpi_xic} (c) shows the 
$M(\Lambda_{c}^{+}K^{-}\pi^{+})$ distribution for the 
$\Sigma_c(2455)^{++}$ signal region
together with the same plot for the $\Sigma_c(2455)^{++}$ sideband region, defined as 
$|M(\Lambda_c^{+}\pi^{+})-(m_{\Sigma_c(2455)^{++}}\pm 15$ MeV/${\it c}$$^{2}$ $) | \: <$ 5 MeV/${\it c}$$^{2}$. 
Clear peaks corresponding to the $\Xi_c(2980)^{+}$, $\Xi_c(3055)^{+}$ and $\Xi_c(3080)^{+}$ are seen. 
To obtain the statistical significance of the $\Xi_c(3055)^{+}$, an UML
fit is applied. PDFs for the $\Xi_c^{\ast +}$ components are represented 
by a Breit-Wigner line-shape convolved
with a Gaussian to account for the invariant-mass resolution ($\sigma_{\rm res}$). Using the signal MC events,
we estimate $\sigma_{\rm res}$ to vary from 1.2 to 1.8 MeV/${\it c}$$^{2}$, depending on the masses of the $\Xi_c^{\ast +}$ states.
The width and mean of the Breit-Wigner functions are treated as free 
parameters. The background PDF, $f_{1}(x)$, is modeled with a threshold function:
\begin{eqnarray}\label{eq_th1}
\lefteqn{f_{1}(x)=1-\exp((x-x_{0})/\delta_{m})(x/x_{0})^a} \nonumber \\
\lefteqn{\:\:\:\:\:\:\:\:\:\:\:\:\:\ \quad{}+b(x/x_{0}-1)\:\:\:(\textrm{if\ } x>x_{0})} \nonumber \\
\lefteqn{f_{1}(x)=0\:\:\:(\textrm{if\ } x<x_{0}),}
\end{eqnarray}
where $a$, $b$, $x_{0}$, and $\delta_{m}$ are free parameters in the fit.

The fit result is shown in Fig. \ref{momdis_lamkpi_xic} (c).
To estimate the statistical significance of the $\Xi_{c}(3055)^{+}$,
we compare the likelihood values for the fits with and without the
$\Xi_{c}(3055)^{+}$ component.
The obtained $-2\ln{(\mathcal{L}_{0}/\mathcal{L})}$ value is 54.7.
By taking into account the change of the number of degrees of freedom 
($ndf$) by the
inclusion of the $\Xi_{c}(3055)^{+}$ component,
the statistical significance of the $\Xi_{c}(3055)^{+}$ becomes 6.8$\sigma$.
The $\chi^{2}/ndf$ of the fit with the $\Xi_{c}(3055)^{+}$ component, 
for the binning of Fig. \ref{momdis_lamkpi_xic} (c), is 54.8/61.

%All the values except for the mass of $\Xi_c(3055)^{+}$ are found to be consistent with PDG values within 2 $\sigma$.

Figure \ref{momdis_lamkpi_xic} (d) shows the 
$M(\Lambda_{c}^{+}K^{-}\pi^{+})$ distribution for the 
$\Sigma_c(2520)^{++}$ selected region
together with the same plot for the $\Sigma_c(2520)^{++}$ 
sideband region, defined as 
$|M(\Lambda_c^{+}\pi^{+})-(m_{\Sigma_c(2520)^{++}}\pm27$ MeV/${\it c}$$^{2}$ $)| \: <$ 12 MeV/${\it c}$$^{2}$.
A clear peak corresponding to the $\Xi_c(3080)^{+}$ is seen, 
while no peak structure is seen in the mass near 3.123 GeV/${\it c}$$^{2}$.
An UML fit is applied to extract the signal yield.
Again, the $\Xi_c^{ \ast +}$ components are represented by
a Breit-Wigner function convolved with a Gaussian. 
For the $\Xi_c(3080)^{+}$ component, the mass and width of the 
Breit-Wigner are
treated as free parameters; while for the $\Xi_c(3123)^{+}$ component, the mass 
and width are fixed to the values obtained in Ref.\cite{Aubert:2007dt}.
The background shape, $f_{2}(x)$, is assumed to be:  
\begin{eqnarray}\label{eq_th2}
\lefteqn{f_{2}(x)=(1-\exp((x-x_{0})^2/\delta_{m}))(x/x_{0})^{a}} \nonumber \\
\lefteqn{\:\:\:\:\:\:\:\:\:\:\:\:\:\: \quad{} +b(x/x_{0}-1)+c((x/x_{0})^{2}-1)\:(\textrm{if\ } x>x_{0})} \nonumber \\
\lefteqn{f_{2}(x)=0\:(\textrm{if\ } x<x_{0}),}
\end{eqnarray}
where  $a$, $b$, $c$, $x_{0}$, and $\delta_{m}$ are fit parameters.
The $\chi^{2}/ndf$ of the fit with the $\Xi_{c}(3123)^{+}$ component
for the binning of Fig. \ref{momdis_lamkpi_xic} (d) is 28.6/42.
The yield of the $\Xi_c(3123)^{+}$ is 8 $\pm$ 22 events, 
which is consistent with zero. 
Hence, a 95$\%$ C.L. upper limit for the production cross section 
is evaluated with the method described in the previous section. 
To directly compare with the BaBar result in Ref.~\cite{Aubert:2007dt},
the upper limit for the product of the cross section and branching fraction of $\Lambda_{c}^{+}$
produced with $x_{p}>0.7$ condition,
\begin{eqnarray*}
\sigma_{{\cal B} \Lambda_c^{+}} &\equiv&  \sigma(e^{+}e^{-}\to \Xi_{c}(3123)^{+} X) \times {\cal B}(\Lambda_c^{+}\to pK^{-}\pi^{+})\\
&=&\frac{N_{{\rm sig}}}{2L(\epsilon_{pK^{-}\pi^{+}}+\epsilon_{pK_{S}} \times {\cal B} _{pK_{S}}/{\cal B} _{pK^{-}\pi^{+}})},
\end{eqnarray*}
is evaluated.
As in Ref.~\cite{Aubert:2007dt}, we assume ${\cal B}(\Xi_c(3123)^{+}\to\Sigma_c(2520)^{++}K^{-})$ is equal to 1. 

To take the uncertainty of the $\Xi_{c}(3123)^{+}$ mass and width from Ref.~\cite{Aubert:2007dt}
into account, we perform a pseudo-experiment test.
The background and $\Xi_{c}(3080)^{+}$ contributions are generated with  
statistics similar to data and based on the fit result. 
The $\Xi_{c}(3123)^{+}$ component is generated with mass and width changed 
by $\pm$1$\sigma$, corresponding to their 
measured uncertainties (3122.9 $\pm$ 1.3 $\pm$ 0.4 MeV/${\it c}$$^{2}$ for the mass and 4.4 $\pm$ 3.4 $\pm$ 1.7 MeV/${\it c}$$^{2}$ for the 
width).
The yield of $\Xi_{c}(3123)^{+}$ is extracted by fitting 
pseudo-experiment data with the procedure used for data.
The ratio of the generated  and extracted yield is regarded as a 
systematic  uncertainty.
Because the error of the width is relatively large, its systematic uncertainty 
contribution is dominant (28$\%$).
All of the systematic errors are summarized in the third column of 
Table \ref{summary_sys}.
The 95$\%$ C.L. upper limit on $\sigma_{{\cal B}  \Lambda_c^{+}}$ 
is 0.17 fb.
As in the case of the $\Xi_{cc}$ cross section, the measurement in Ref.~\cite{Aubert:2007dt} 
does not introduce a  factor of two for the $\sigma_{{\cal B} \Lambda_c^{+}}$ calculation.
Therefore, we should double our measurement, which results in 
0.34 fb, when comparing with BaBar's result.
The value is much smaller than that quoted in Ref.~\cite{Aubert:2007dt}
(1.6 $\pm$ 0.6 $\pm$ 0.2 fb).

The systematic uncertainties of the masses and widths of the $\Xi_{c}^{\ast +}$
and stability of the statistical significance of the $\Xi_{c}(3055)^{+}$ are studied 
by the following fitting configurations.
The systematic uncertainties due to the signal PDF are studied by varying 
$\sigma_{{\rm res}}$ by 5$\%$.
The systematic uncertainties due to possible interference between 
the $\Xi_{c}(3055)^{+}$ and $\Xi_{c}(3080)^{+}$ are studied 
by fitting the distribution with an additional phase parameter 
between the two Breit-Wigner amplitudes.
The systematic uncertainty due to the background shape
is studied by fitting the mass spectra with a second-order polynomial as a background 
PDF in the range of 3.005-3.200 GeV/${\it c}$$^{2}$. 
%The statistical significance of the $\Xi_{c}(3055)^{+}$ do not fall below 6.6$\sigma$
%in any of these fitting configurations.
In none of these fitting configurations does the statistical significance of the 
$\Xi_{c}(3055)^{+}$ fall below 6.6$\sigma$.
We apply cut conditions of $x_{p}>0.6$ and $x_{p}>0.8$ instead of 
$x_{p}>0.7$ and re-extract the masses and widths of the $\Xi_{c}^{\ast +}$ states.
The differences from the default cut condition are regarded as 
systematic uncertainties.
The measured masses, widths and yields of the three $\Xi_{c}^{\ast +}$ states 
are summarized in Table \ref{summary_sigmac}.
All of these measurements are consistent with previous Belle measurement~\cite{Chistov:2006zj}
within 2.5$\sigma$ and with the BaBar measurement~\cite{Aubert:2007dt} within 2.0$\sigma$.

%\begin{table}[htbp]
%  \begin{center}
%  \caption{The measured masses and widths of the three $\Xi_c^{\ast +}$ states.
%           First error is statistical and second one is systematic.}
%    \begin{tabular}{|l|l|l|l|} \hline
%                                             & Mass (MeV/${\it c}$$^{2}$)         & Width (MeV/${\it c}$$^{2}$)       & Yield \\ \hline
%      $\Xi_c(2980)^{+}$                      & 2974.9 $\pm$ 1.5 $\pm$ 2.1 & 14.8 $\pm$ 2.5 $\pm$ 4.1  & 244 $\pm$ 39\\ \hline 
%      $\Xi_c(3055)^{+}$                      & 3058.1 $\pm$ 1.0 $\pm$ 2.1 &  9.7 $\pm$ 3.4 $\pm$ 3.3  & 199$\pm$ 46 \\ \hline 
%      $\Xi_c(3080)^{+}$($\Sigma_{c}$)        & 3077.9 $\pm$ 0.4 $\pm$ 0.7 &  3.2 $\pm$ 1.3 $\pm$ 1.3  & 185 $\pm$ 31\\ \hline 
%      $\Xi_c(3080)^{+}$($\Sigma_{c}^{\ast}$) & 3076.9 $\pm$ 0.3 $\pm$ 0.2 &  2.4 $\pm$ 0.9 $\pm$ 1.6  & 210 $\pm$ 30\\ \hline 
%
%    \end{tabular}
%  \end{center}
%  \label{summary_masswidthsigmac}
%\end{table}

\begin{center}
  \begin{table*}[htbp]
    \caption{The measured masses and widths of the three $\Xi_c^{\ast +}$ states.
      The first error is statistical and second is systematic.}
    \begin{tabular}{l|lll} \hline \hline
      Particle                            & Mass (MeV/${\it c}$$^{2}$)         & Width (MeV/${\it c}$$^{2}$)       & Yield \\ \hline 
      $\Xi_c(2980)^{+}$                      & 2974.9 $\pm$ 1.5 $\pm$ 2.1 & 14.8 $\pm$ 2.5 $\pm$ 4.1  & 244 $\pm$ 39\\ %\hline 
      $\Xi_c(3055)^{+}$                      & 3058.1 $\pm$ 1.0 $\pm$ 2.1 &  9.7 $\pm$ 3.4 $\pm$ 3.3  & 199 $\pm$ 46 \\ %\hline 
      $\Xi_c(3080)^{+}$($\Sigma_{c}$)        & 3077.9 $\pm$ 0.4 $\pm$ 0.7 &  3.2 $\pm$ 1.3 $\pm$ 1.3  & 185 $\pm$ 31\\ %\hline 
      $\Xi_c(3080)^{+}$($\Sigma_{c}^{\ast}$) & 3076.9 $\pm$ 0.3 $\pm$ 0.2 &  2.4 $\pm$ 0.9 $\pm$ 1.6  & 210 $\pm$ 30\\ \hline \hline
    \end{tabular}
    \label{summary_sigmac}
  \end{table*}
\end{center}

\section{Final state with $\Xi_{c}^{0}$}\label{section_xic}
In this section, the analysis of the final state with the $\Xi_{c}^{0}$ 
is described.
The reconstruction of the $\Xi_{c}^{0}$ is presented first, followed by the 
analysis of the $\Xi_{c}(2645)^{+}$.
Finally, a search for $\Xi_{cc}^{+(+)}$ decaying into 
the $\Xi_{c}^{0}\pi^{+}(\pi^{+})$ final state is described.

\subsection{Reconstruction of $\Xi_{c}^{0}$}
The $\Xi_{c}^{0}$ is reconstructed in three decay modes: 
$\Xi^{-}\pi^{+}$, $\Lambda K^{-} \pi^{+}$ and $pK^{-}K^{-}\pi^{+}$.
The $\Lambda$ is reconstructed from its decay into $p \pi^{-}$.
%, with the
%proton from the decay of $\Lambda$ is required to have 
%likelihood ratios $\mathcal{L}(p:\pi)$ and $\mathcal{L}(p:K)$ to be greater than 0.6.
The proton and $\pi^{-}$ tracks for $\Lambda$ candidates are fitted to a common vertex.
The fitting result is used to suppress misreconstructed $\Lambda$ candidates and 
to perform the subsequent vertex fit for the $\Xi^{-}\to \Lambda \pi^{-}$ or $\Xi_{c}^{0} \to \Lambda K^{-} \pi^{+}$.
The invariant mass of the $\Lambda$ candidate is required to be within 
3 MeV/${\it c}$$^{2}$ of the nominal $\Lambda$ mass, which corresponds to 
approximately 3$\sigma$ of the mass resolution.
The selection based on their decay vertex information is also applied~\cite{Lambdaselection}. 
The $\Xi^{-}$ is reconstructed from its decay into $\Lambda \pi^{-}$.
The $\Lambda$ and $\pi^{-}$ tracks for $\Xi^{-}$ candidates are fitted to a common vertex.
The fitting result is used to clean up the $\Xi^{-}$ candidates and 
in the common vertex fit for the $\Xi_{c}^{0} \to \Xi^{-} \pi^{+}$.
The closest distance of the $\Lambda$ and $\pi^{-}$ along the $z$-direction 
is required to be less than 3 mm.
We require $\cos \theta >0.95$, where $\theta$ is the angle between 
the momentum vector of the $\Xi^{-}$
and the vector between the IP and the $\Xi^{-}$ decay vertex.
The $\chi^{2}$ of the common-vertex fit of the $\Lambda \pi^{-}$ is 
required to be less than 50.
The invariant mass of a $\Xi^{-}$ candidate is required to be within 
4 MeV/${\it c}$$^{2}$ of the nominal $\Xi^{-}$ mass, which corresponds to 
approximately 3$\sigma$ of the mass resolution.
The daughter particles of the $\Xi_{c}^{0}$ are fitted to a common vertex.
The $\Xi_{c}^{0}$ candidates are selected by requiring invariant masses 
of the daughter particles with common vertex fit to be within
12, 7 and 7 MeV/${\it c}$$^{2}$ of the nominal $\Xi_{c}^{0}$ mass 
for the $\Xi^{-}\pi^{+}$, $\Lambda K^{-}\pi^{+}$
and $p K^{-} K^{-} \pi^{+}$ decay modes, respectively, which correspond to approximately 1.5$\sigma$
of the mass resolution. 
The $\chi^{2}$ value of the common vertex fit for the $\Xi_{c}^{0}$ is required to be less than 50.
The mass constraint fit to the $\Xi_{c}^{0}$ mass is performed.

We optimize the selection criteria for $x_{p}$ in the  
$\Xi_{c}^{0}\pi^{+}(\pi^{+})$ system with the method
described in \ref{section_xicc_lambdac}, again assuming that the 
$x_{p}$ spectrum for the $\Xi_{cc}$ is the same as 
that for the $\Lambda_{c}^{+}$. We require $0.45<x_{p}<1.0$ independent 
of the $\Xi_{cc}$ mass and the $\Xi_{c}^{0}$ decay mode. The same cut 
is applied for the analysis of the $\Xi_{c}(2645)^{+}$.

\subsection{Study of the $\Xi_c^{+}(2645)$}\label{subsection_xic2645}
Unlike the $\Lambda_{c}^{+}$ study, the signal-to-background ratio of the 
$\Xi_{c}^{0}$ largely depends on the 
decay modes. Therefore, to improve our sensitivity for the $\Xi_{cc}$, 
we perform a simultaneous fit to the mass spectra 
for the three $\Xi_{c}^{0}$ decays with fixed relative signal ratios. 
We use the relative yields of the 
$\Xi_{c}(2645)^{+} \to \Xi_{c}^{0}\pi^{+}$ measured for the 
$\Xi_{c}^{0}$ decay modes 
to estimate a relative signal yield of the $\Xi_{cc}$.
The relative signal yields of the $\Xi_{cc}$ ($N^{i}_{\Xi_{cc}}$) 
in a given $\Xi_c^{0}$ decay channel can be written as 
\begin{eqnarray}\label{equ_relative}
N^{i}_{\Xi_{cc}}=N^{i}_{\Xi_c(2645)^{+}}\frac{\epsilon^{i}_{\Xi_{cc}}}{\epsilon^{i}_{\Xi_{c}(2645)^{+}}},
\end{eqnarray}
where $N^{i}_{\Xi_c(2645)^{+}}$ is the $\Xi_c(2645)^{+}$ yield, 
$\epsilon^{i}_{\Xi_{cc}}$
is the reconstruction efficiency of the $\Xi_{cc}$,
and $\epsilon^{i}_{\Xi_{c}(2645)^{+}}$
is the reconstruction efficiency of the $\Xi_c(2645)^{+}$. 
Both efficiencies include the secondary branching fractions for 
$\Lambda \to p \pi^{-}$ of (63.9 $\pm$ 0.5)$\%$ and 
$\Xi^{-} \to \Lambda \pi^{-}$ of (99.887 $\pm$ 0.035)$\%$.
The index $i$ denotes the decay mode of the $\Xi_c^{0}$.

\begin{figure*}[htbp]
  \begin{center}
    \includegraphics[scale=0.29]{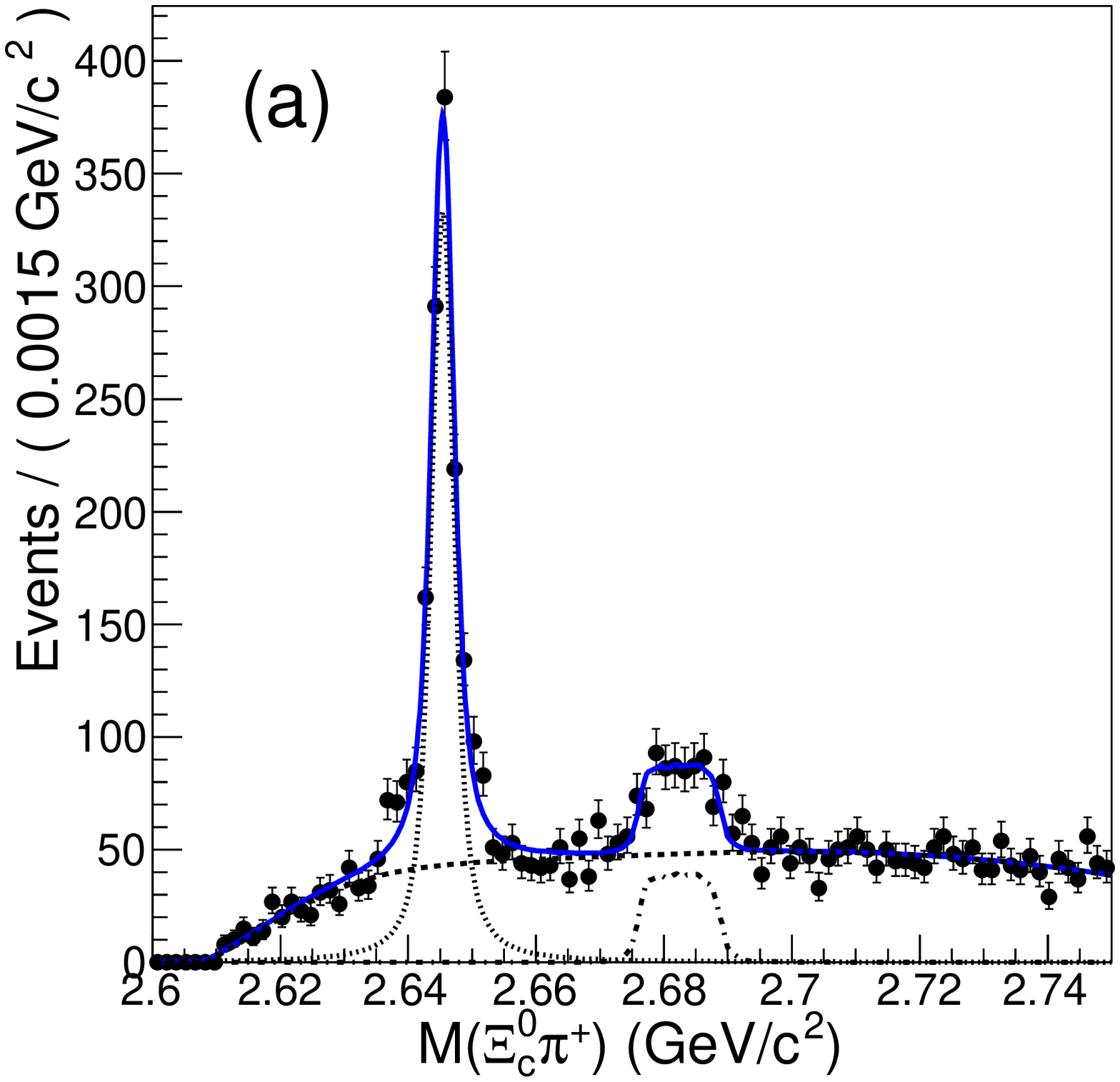}
    \includegraphics[scale=0.29]{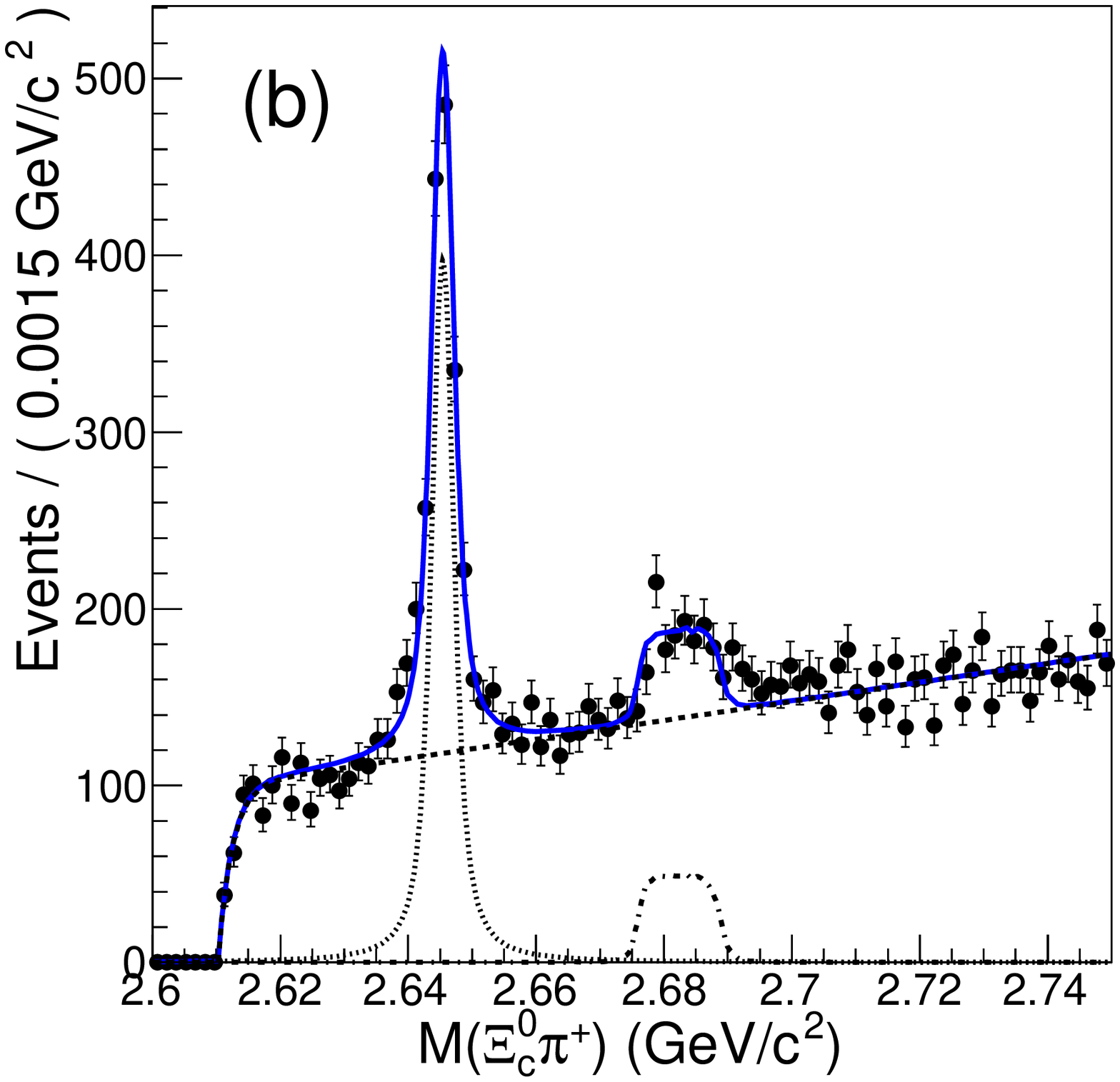}
    \includegraphics[scale=0.29]{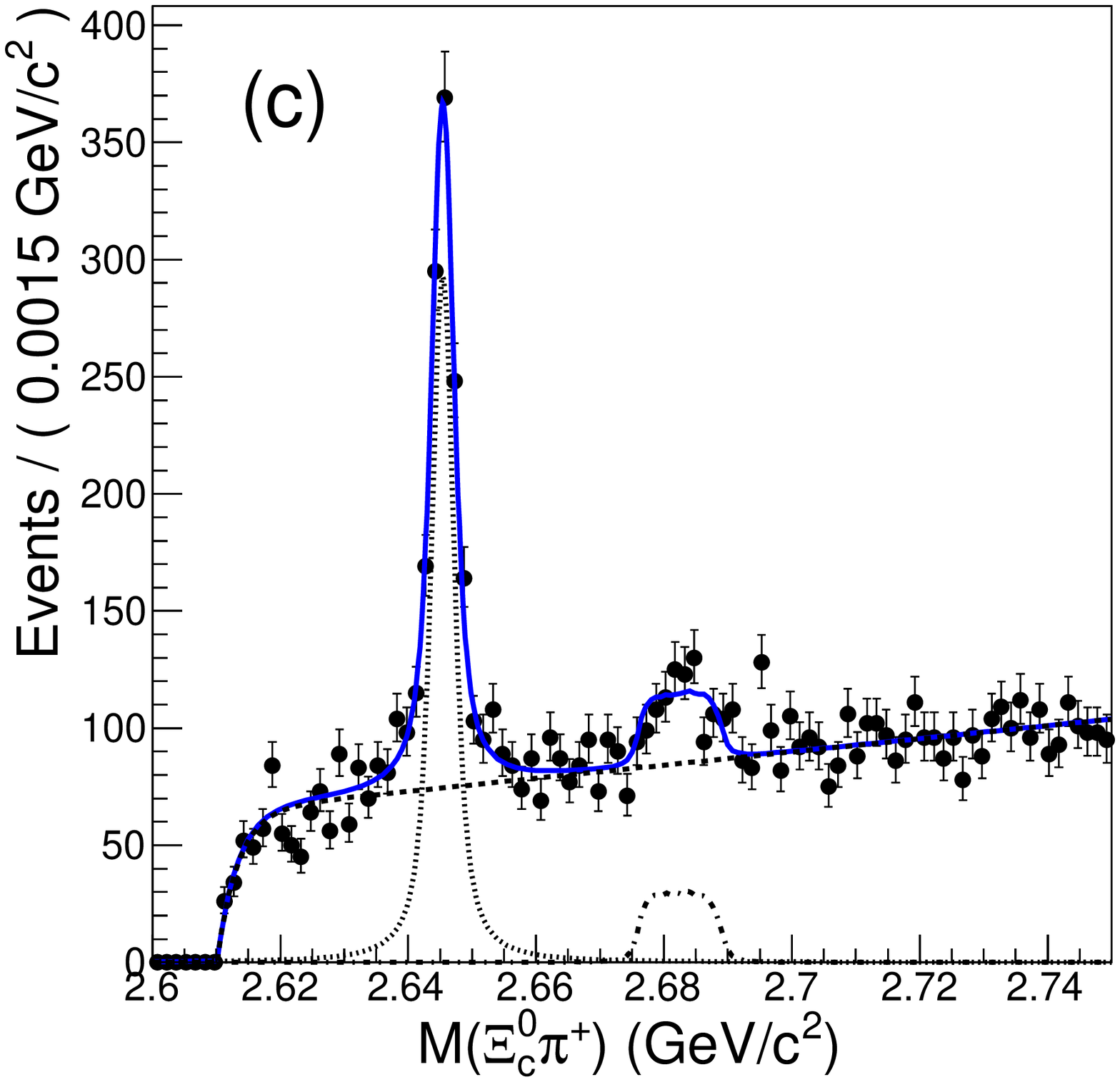}
    \caption{$M(\Xi_c^{0}\pi^{+})$ distributions below the $\Xi_{cc}$ 
              search region for (a) $\Xi_c^{0}\to\Xi^{-}\pi^{+}$, 
             (b) $\Xi_c^{0}\to \Lambda K^{-}\pi^{+}$, 
             (c) $\Xi_c^{0}\to pK^{-}K^{-}\pi^{+}$. The solid lines show the fit result.
             The dashed, dotted, and dash-dotted lines show the contributions from background, $\Xi_c(2645)^{+}$, and
	     $\Xi_c(2790)^{+}$, respectively.
             }
    \label{mgpi_2645}
  \end{center}
\end{figure*}

Figure \ref{mgpi_2645} shows the 
$M(\Xi_c^{0}\pi^{+}$) distribution for each $\Xi_c^{0}$ decay mode below the
$\Xi_{cc}$ search region.
Clear peaks corresponding to the $\Xi_c(2645)^{+}$ are seen in 
all decay modes. The bump structures near 2.68 GeV/${\it c}$$^{2}$ 
originate from the process 
$\Xi_{c}(2790)^{+} \to \Xi_{c}'^{0}\pi^{+} \to \Xi_{c}^{0} \gamma \pi^{+} $ 
with a $\gamma$ missing in the reconstruction.
The simultaneous UML fit is applied 
to extract the relative yields and the width of the $\Xi_c(2645)^{+}$.
The $\Xi_c(2645)^{+}$ signal is represented by a Breit-Wigner function 
convolved with a Gaussian whose width
corresponds to the mass resolution $\sigma_{{\rm res}}$.
The value of $\sigma_{{\rm res}}$ is 1.05 MeV/${\it c}$$^{2}$, independent
of the decay modes of the $\Xi_{c}^{0}$.
The PDF of the $\Xi_{c}(2790)^{+}$ reflection is modeled using MC.
$f_{2}(x)$ in Eq. \ref{eq_th2} is used as the background PDF for 
the $\Xi_c^{0} \to \Xi^{-}\pi^{+}$ decay mode
whereas $f_{1}(x)$ in Eq. \ref{eq_th1} is used for the 
$\Xi_c^{0} \to \Lambda K^{-} \pi^{+}$ and $\Xi_c^{0} \to p K^{-}K^{-}\pi^{+}$ 
decay modes.
The width and mass of the $\Xi_c(2645)^{+}$ are constrained to be the same 
for the three decay modes.
The yield of the $\Xi_c(2645)^{+}$ is 1298 $\pm$ 51, 1444 $\pm$ 58 and 974 $\pm$ 47  
for the $\Xi^{-}\pi^{+}$, $\Lambda K^{-}\pi^{+}$ and $pK^{-}K^{-}\pi^{+}$ 
decay mode, respectively. 
The mass and width are obtained to be 2645.4 $\pm$ 0.1 MeV/${\it c}$$^{2}$ 
and 2.6 $\pm$ 0.2 MeV/${\it c}$$^{2}$, respectively.
The $\chi^{2}/ndf$ of the fit for the binning in Fig. \ref{mgpi_2645}  
is 296/276. 

To check the consistency of the width measurement 
between the  $\Xi_{c}^{0}$ decay modes, we fit 
the three mass spectra separately.
The measured widths are found to be consistent between the three decay modes: 2.9 $\pm$ 0.3 MeV/${\it c}$$^{2}$, 
2.6 $\pm$ 0.3 MeV/${\it c}$$^{2}$ and 2.5 $\pm$ 0.3 MeV/${\it c}$$^{2}$ 
for $\Xi_c^{0}\to$ $\Xi^{-}\pi^{+}$, $\Lambda K^{-}\pi^{+}$ and $pK^{-}K^{-}\pi^{+}$, respectively.
%Masses are obtained to be 2645.4 $\pm$ 0.1 MeV/c$^{2}$, 2645.3 $\pm$ 0.1 MeV/c$^{2}$, and 2645.5 $\pm$ 0.1 MeV/c$^{2}$ 
%for the $\Xi_c^{0}$ decay mode of $\Xi^{-}\pi^{+}$, $\Lambda K^{-}\pi^{+}$, and $pK^{-}K^{-}\pi^{+}$, respectively.
The measured width is found to be consistent for the three decay modes.
The systematic  uncertainty of the width measurement due to the fit procedure is 
studied with pseudo-experiment events samples:
the $\Xi_c(2645)^{+}$ component is generated according to the signal MC sample with the natural width of 
2.6 MeV/${\it c}$$^{2}$ by signal MC, while 
contributions from background and $\Xi_{c}(2790)^{+}$ reflection
are generated based on the fit result with the real data. The statistics 
of the pseudo-experiment samples are  the same as for 
those of data. The width of the $\Xi_c(2645)^{+}$ is extracted from 
simultaneous fits to pseudo-experiment samples,
and its mean value is obtained to be 2.75 $\pm$ 0.03 MeV/${\it c}$$^{2}$, 
which is higher than the input value by 0.15 MeV/${\it c}$$^{2}$. 
The difference is included as the systematic  uncertainty
from the fit procedure.
The systematic  uncertainty due to the background shape is studied 
by fitting the data with a second-order polynomial function 
for the alternative background shape.
The fit region is restricted to 2.62-2.75 GeV/${\it c}$$^{2}$. 
The width is obtained to be 
2.9 $\pm$ 0.2 MeV/${\it c}$$^{2}$, which is 0.3 MeV/${\it c}$$^{2}$ 
higher than the default measurement.
This deviation is included as a systematic  uncertainty.
To check the systematic uncertainty due to the mass resolution,
we evaluate the ratio of the resolution of the $\Xi_{c}^{0}$ in the data 
and MC, $\frac{\sigma_{\rm data}}{\sigma_{\rm mc}}$,
where $\sigma_{\rm data}$ is the resolution of $\Xi_c^{0}$ for data and $\sigma_{\rm mc}$
is that for MC. An additional cut of $2.64$ GeV/${\it c}$$^{2}<M(\Xi_c^{0}\pi^{+})<$ 2.65 GeV/${\it c}$$^{2}$
is applied to select the $\Xi_c(2645)^{+}$ region only.
$\frac{\sigma_{\rm data}}{\sigma_{\rm mc}}$ is 0.96 $\pm$ 0.03, 
1.03 $\pm$ 0.07 and 1.07 $\pm$ 0.03 for the $\Xi^{-}\pi^{+}$, 
$\Lambda K^{-}\pi^{+}$
and $pK^{-}K^{-}\pi^{+}$ decay modes, respectively. We assign the 
systematic  uncertainty conservatively by increasing 
the mass resolution by 7$\%$, which is the largest deviation of the 
$\frac{\sigma_{\rm data}}{\sigma_{\rm mc}}$
from unity, for all the decay modes.
The resulting width is 2.5 $\pm$ 0.2 MeV/${\it c}$$^{2}$. 
The difference of 0.1 MeV/${\it c}$$^{2}$ with respect to the default
measurement is included as a systematic  uncertainty. 
We use alternative $x_{p}$ range of $0.35<x_{p}<1.0$ and $0.55<x_{p}<1.0$ and extract the 
width of 2.6 $\pm0.2$ MeV/${\it c}$$^{2}$ for the former case 
and 2.8 $\pm 0.2$ MeV/${\it c}$$^{2}$ for the latter.
The largest difference of 0.2 MeV/${\it c}$$^{2}$ with respect to the default
measurement is included as a systematic  uncertainty. 
By adding all the systematic  uncertainties in quadrature,
the total systematic uncertainty for the width measurement 
is estimated to be 0.4 MeV/${\it c}$$^{2}$.

%The width measurement for the each $\Xi_{c}^{0}$ decay mode is summarized in Table \ref{summary_width_xic2645}.

%\begin{table*}[htbp]
%  \begin{center}
%    \caption{Mass and width of the $\Xi_{c}(2645)^{+}$ and the $\chi^{2}/ndf$ of the various fit variants.}
%    \label{summary_width_xic2645}
%    \begin{tabular}{l|lll} \hline \hline
%     Decay mode  &   Width (MeV/c$^{2}$) &$\chi^{2}/ndf$  \\ \hline
%     $\Xi^{-}\pi^{+}$        & 2.9 $\pm$ 0.3 & 103/91 \\ 
%     $\Lambda K^{-}\pi^{+}$  & 2.6 $\pm$ 0.3 & 103/91 \\ 
%     $pK^{-}K^{-}\pi^{+}$    & 2.5 $\pm$ 0.3 & 91/90  \\ 
%     Simultaneous fit        & 2.6 $\pm$ 0.2 $\pm$ 0.4 & 296/276  \\ \hline \hline
%     PDG                    & 2645.9$^{+0.5}_{-0.6}$ & $<$3.1 &  \\ \hline
%    \end{tabular}
%  \end{center}
%\end{table*}

%\begin{table*}[htbp]
%  \begin{center}
%    \begin{tabular}{l|l} \hline \hline
%      Source          & Error (MeV/c$^{2}$) \\ \hline
%     Input/output     & 0.15 \\ 
%     Background shape & 0.3  \\ 
%     $\sigma_{{\rm res}}$   & 0.1  \\ 
%     Total            & 0.4  \\ \hline
%    \end{tabular}
%  \end{center}
%  \caption{Systematic errors for the width measurement of $\Xi_c(2645)^{+}$.}
%  \label{summary_sys_width}
%\end{table*}

\subsection{Search for doubly charmed baryons in the 
$\Xi_{c}^{0}\pi^{+}(\pi^{+})$ final state}
\begin{figure*}[htbp]
  \begin{center}
    \includegraphics[scale=0.25]{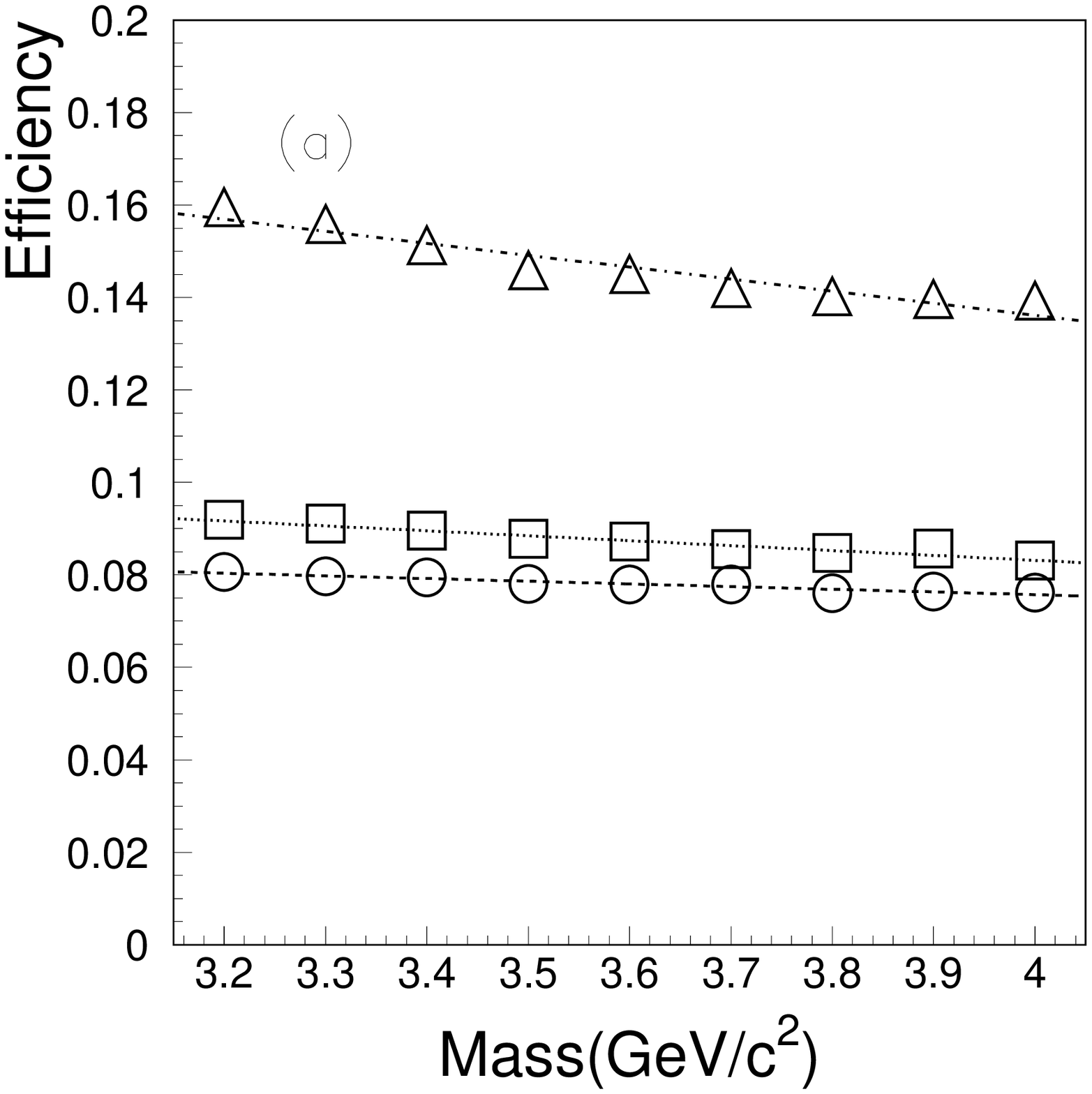}
    \includegraphics[scale=0.25]{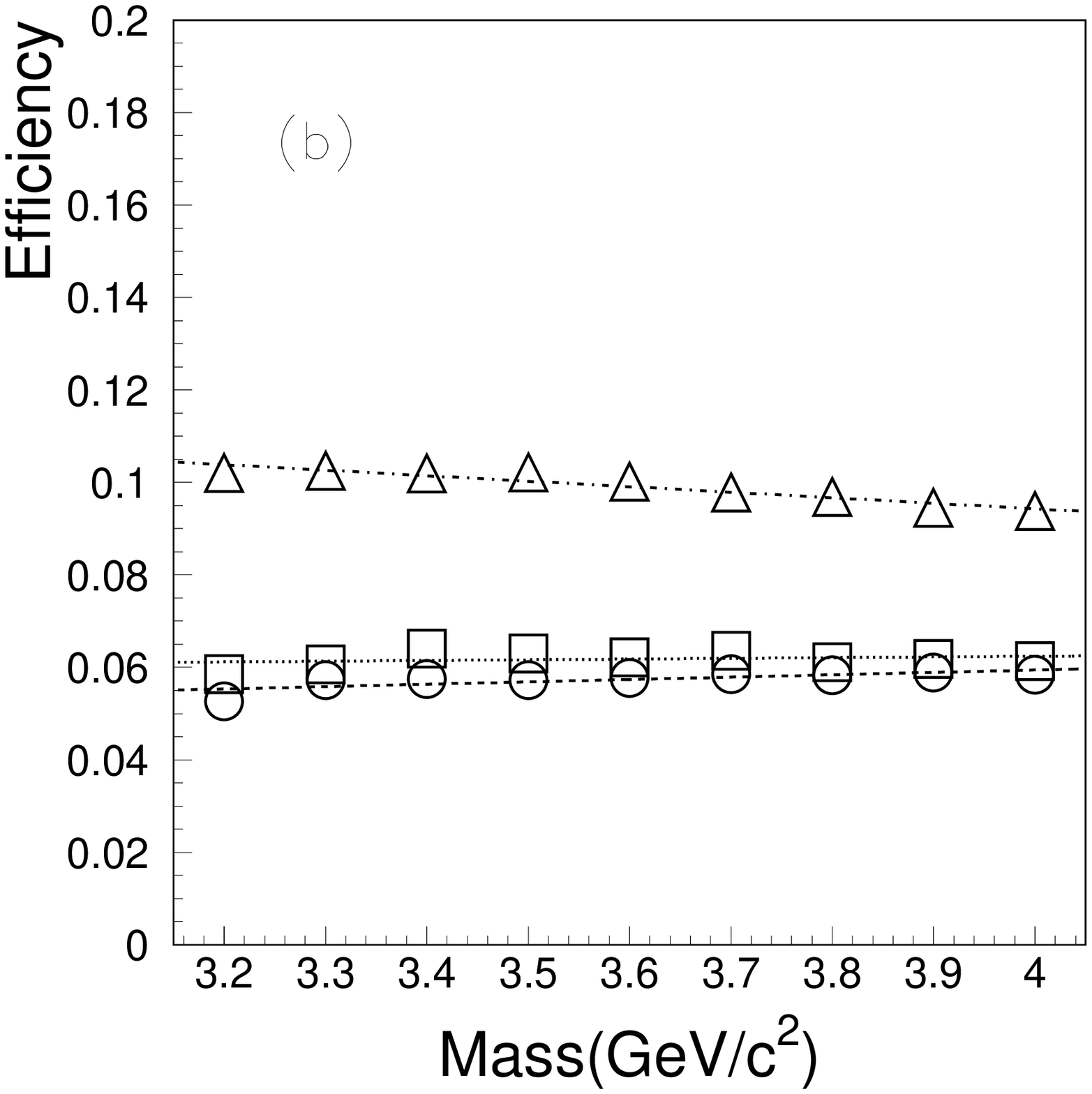}
    \caption{Reconstruction efficiencies for the $\Xi_{cc}$ as a 
function of the $\Xi_{cc}$ mass for 
            (a) $\Xi_{cc}^{+}$, (b) $\Xi_{cc}^{++}$. Circles, square and 
triangle points are for 
             $\Xi_{c}^{0}\to$ $\Xi^{-} \pi^{+}$, $\Lambda K^{-} \pi^{+}$ and $pK^{-}K^{-}\pi^{+}$, respectively.
             The lines are the result of the fit with a linear function.}
    \label{eff_guzaicpi}
  \end{center}
\end{figure*}

To obtain the relative yields of the $\Xi_{cc}$, $\epsilon^{i}_{\Xi_{cc}}$ 
and $\epsilon^{i}_{\Xi_{c}(2645^{+})}$ are
evaluated using MC, and the efficiency $\epsilon^{i}_{\Xi_{cc}}$ is obtained 
as a function of the $\Xi_{cc}$ mass.
$N^{i}_{\Xi_c(2645)^{+}}$ and $\epsilon^{i}_{\Xi_{c}(2645^{+})}$ are 
summarized in Table \ref{summary_nxi2645},
while $\epsilon^{i}_{\Xi_{cc}^{+}}$ as a function of $\Xi_{cc}$ mass is shown 
in Fig.~\ref{eff_guzaicpi}.
As an example, the relative yield ratio of the $\Xi_{cc}^{+}$ with a mass 
of 3.6 GeV/${\it c}$$^{2}$ is
$N^{\Xi^{-}\pi^{+}}_{\Xi_{cc}}$:$N^{\Lambda K^{-} \pi^{+}}_{\Xi_{cc}}$:$N^{p K^{-} K^{-} \pi^{+}}_{\Xi_{cc}}$=1:1.15:0.84.

\begin{table*}[htbp]
  \begin{center}
  \caption{$\Xi_{c}(2645)^{+}$ yields and efficiencies used for estimation of
the relative $\Xi_{cc}$ yields}
  \label{summary_nxi2645}
    \begin{tabular*}{8cm}{@{\extracolsep{\fill}} c| c c c} \hline \hline
%   \begin{tabular}{c| c c c} \hline \hline
     Decay mode & $N_{\Xi_c(2645)^{+}}$ & $\epsilon_{\Xi_{c}(2645)^{+}}$ &   \\  \hline
     $\Xi^{-}\pi^{+}$       & 1298 $\pm$ 51 & 0.0748 $\pm$ 0.0002   \\ 
     $\Lambda K^{-}\pi^{+}$ & 1444 $\pm$ 58 & 0.0977 $\pm$ 0.0003   \\ 
     $pK^{-}K^{-}\pi^{+}$   & 974  $\pm$ 47 & 0.1920 $\pm$ 0.0005   \\ \hline \hline
    \end{tabular*}
  \end{center}
\end{table*}

%The ratio of the number of the background events under the $\Xi_{c}(2645)^{+}$ peak for each decay mode ($N_{\rm bg}^{i}$)
%is $N_{\rm bg}^{\Xi^{-}\pi^{+}}$:$N_{\rm bg}^{\Lambda K^{-}\pi^{+}}$:$N_{\rm bg}^{pK^{-}K^{-}\pi^{+}}$=1.0:3.0:1.4.
%The statistical sensitivity of the $\Xi_{cc}$ is proportional to the $N_{\Xi_{cc}^{i}}/\sqrt{N_{\rm bg}^{i}+N_{\Xi_{cc}^{i}}}$.
%Assuming the $N_{\Xi_{cc}^{i}}$ is much smaller than the $N_{\rm bg}^{i}$
%and neglecting the $\Xi_{cc}$ mass dependence of the $\epsilon^{i}_{\Xi_{cc}^{+}}$ and the ratio of the $N_{\rm bg}^{i}$,
%ratio of the sensitivity is estimated to be 1.0:0.64:0.63. By adding them quadratically, the statistical sensitivity 
%is estimated to be increased by 34 $\%$ when compared with using $\Xi^{-}\pi^{+}$ decay mode only, which 
%corresponds to about 80 $\%$ increase of the statistics.

\begin{figure*}[htbp]
  \begin{center}
    \includegraphics[scale=0.25]{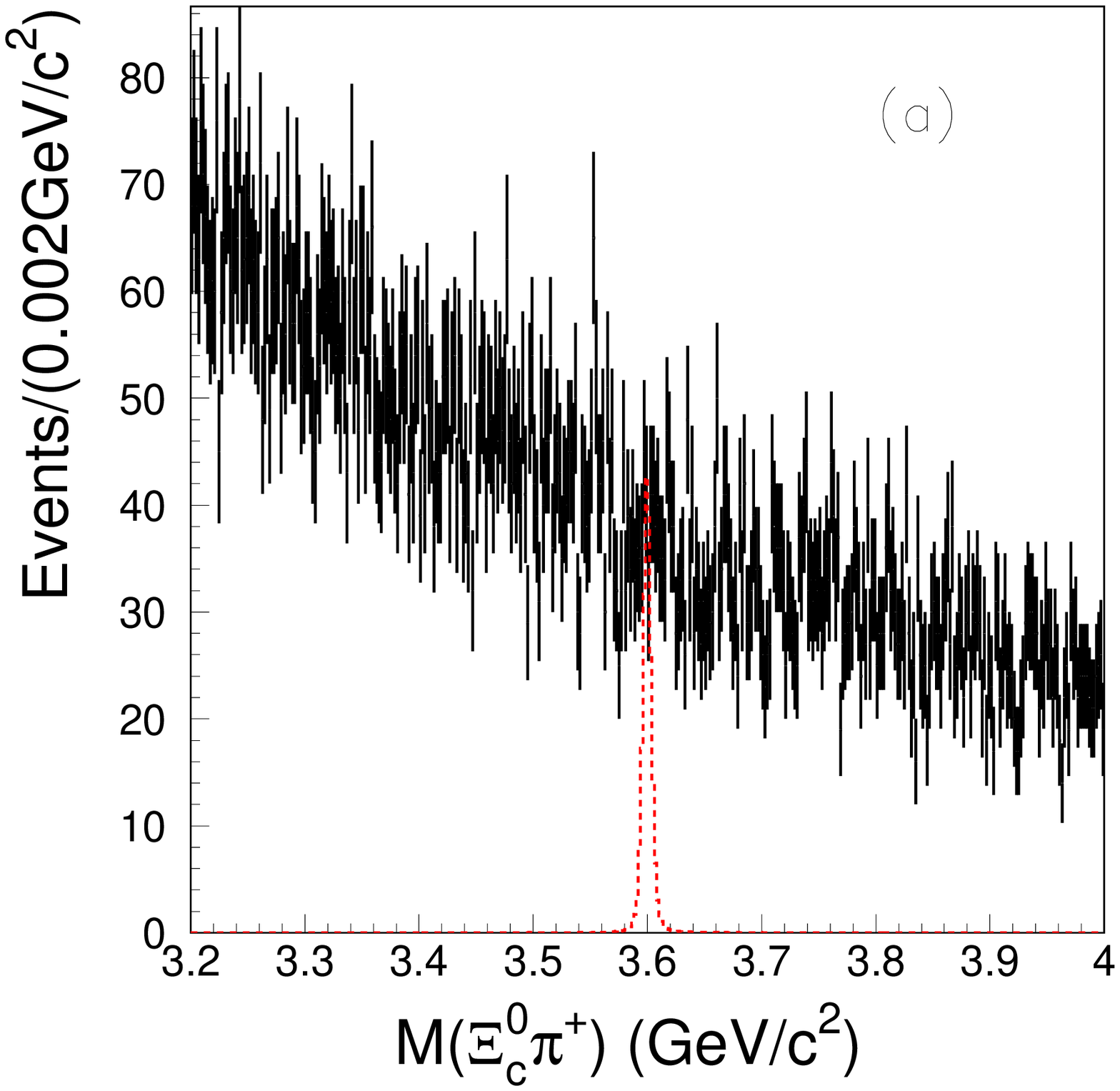}
    \includegraphics[scale=0.25]{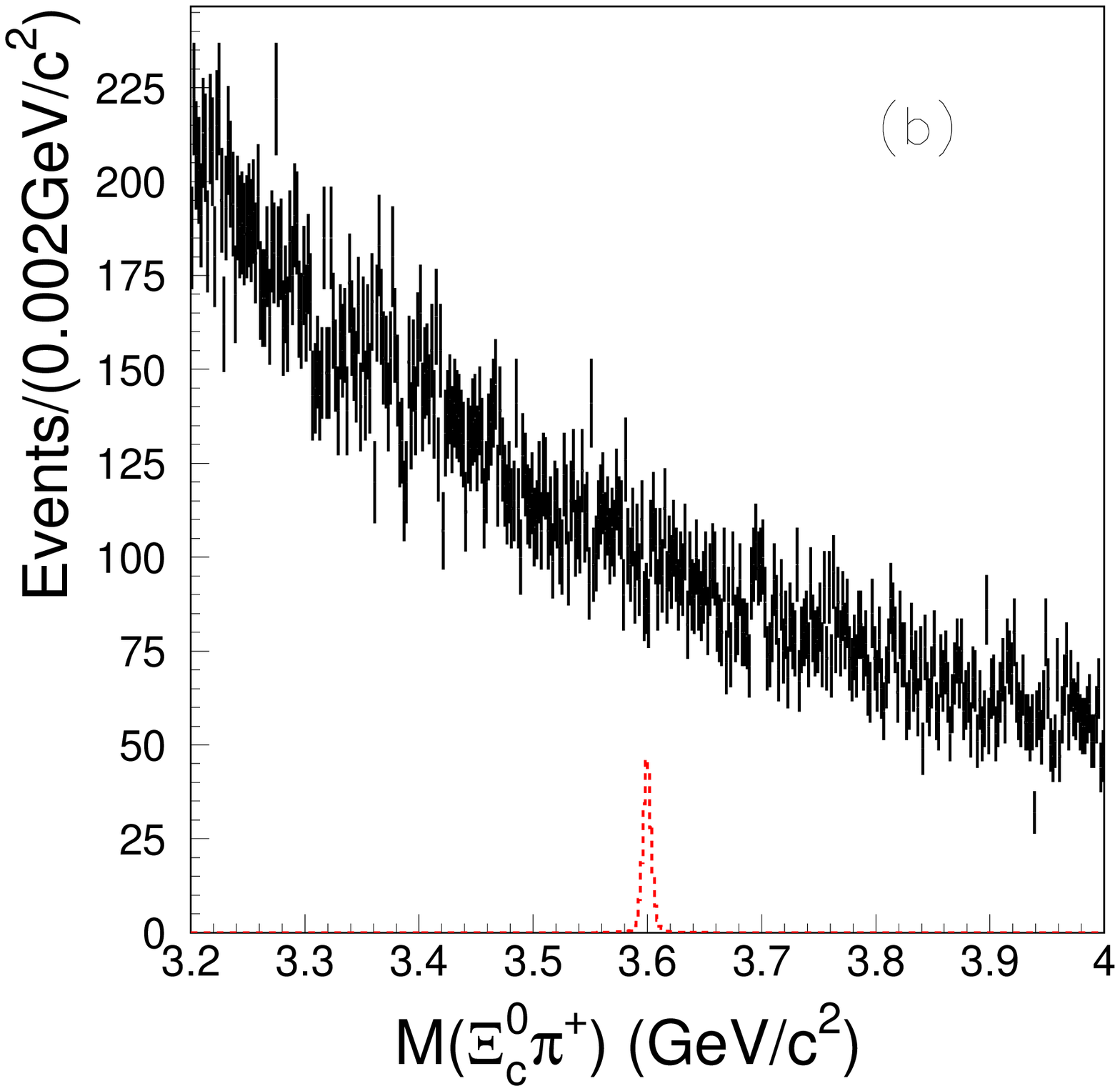}

    \includegraphics[scale=0.25]{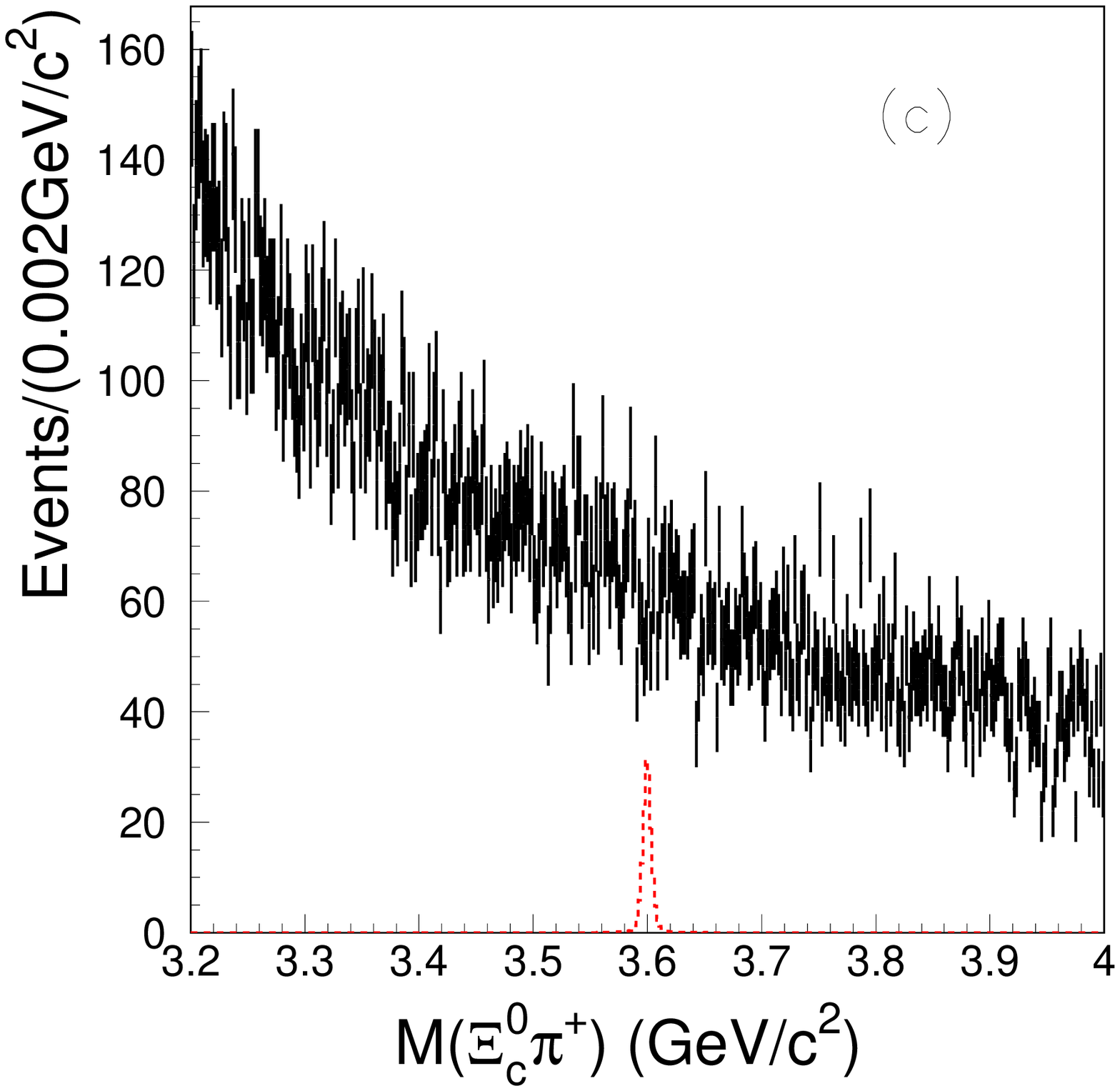}
    \includegraphics[scale=0.25]{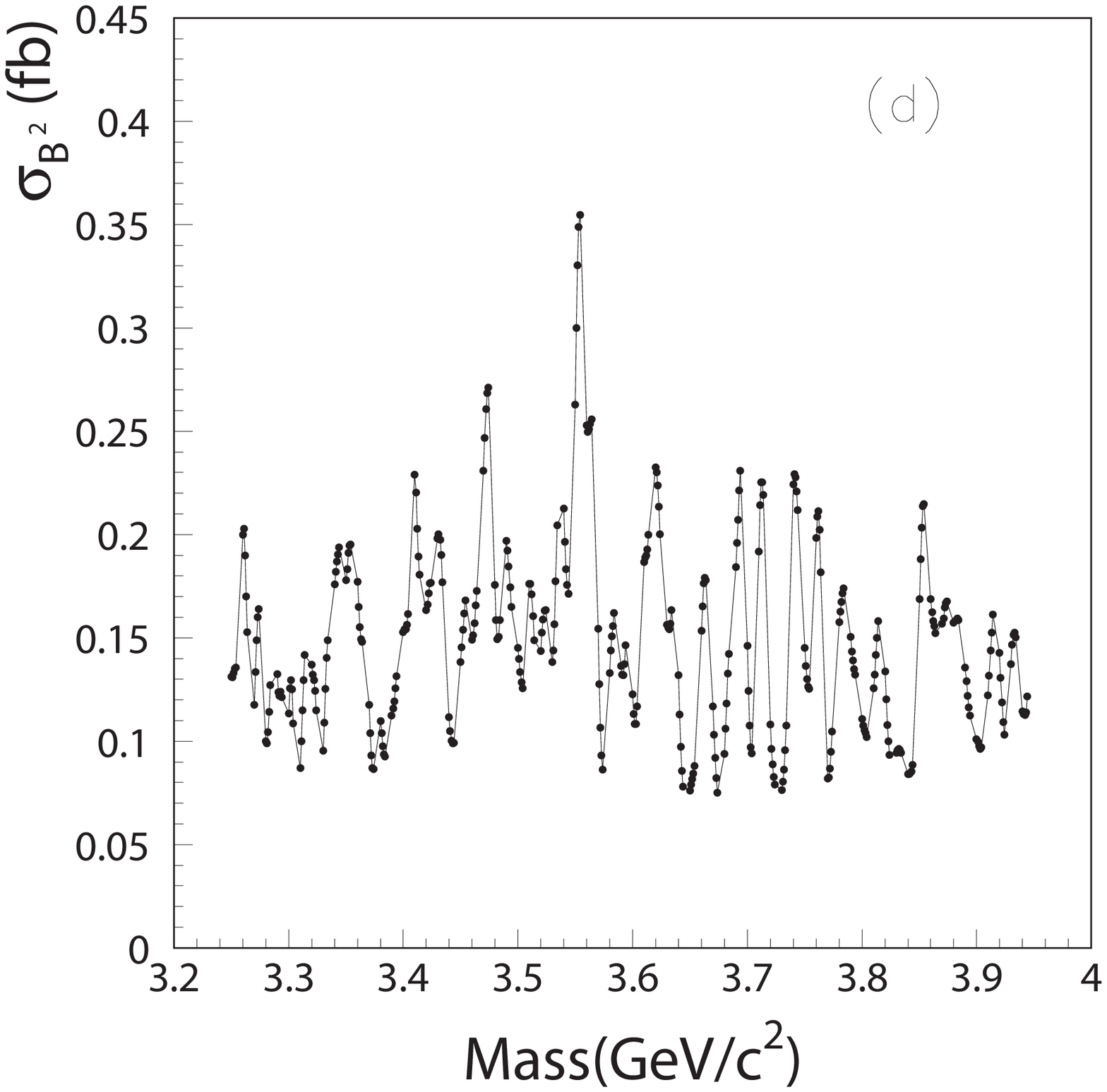}
    \caption{(a)-(c):$M(\Xi_{c}^{0}\pi^{+})$ distribution in the 
$\Xi_{cc}^{+}$ search region
             for $\Xi_{c}^{0}\to$ (a) $\Xi^{-}\pi^{+}$,
                                  (b) $\Lambda K^{-}\pi^{+}$, 
                                  (c) $pK^{-}K^{-}\pi^{+}$.
             The vertical error bars are from data. The dashed histograms are from signal MC.
            (d): 95$\%$ C.L. upper limit of the $\sigma_{{\cal B}^{2}}$ 
for $\Xi_{cc}^{+}$
                as a function of the mass with a 
1 MeV/${\it c}$$^{2}$ step.}
    \label{open_guzaic+pi_momcut}
  \end{center}
\end{figure*}

\begin{figure*}[htbp]
  \begin{center}
    \includegraphics[scale=0.25]{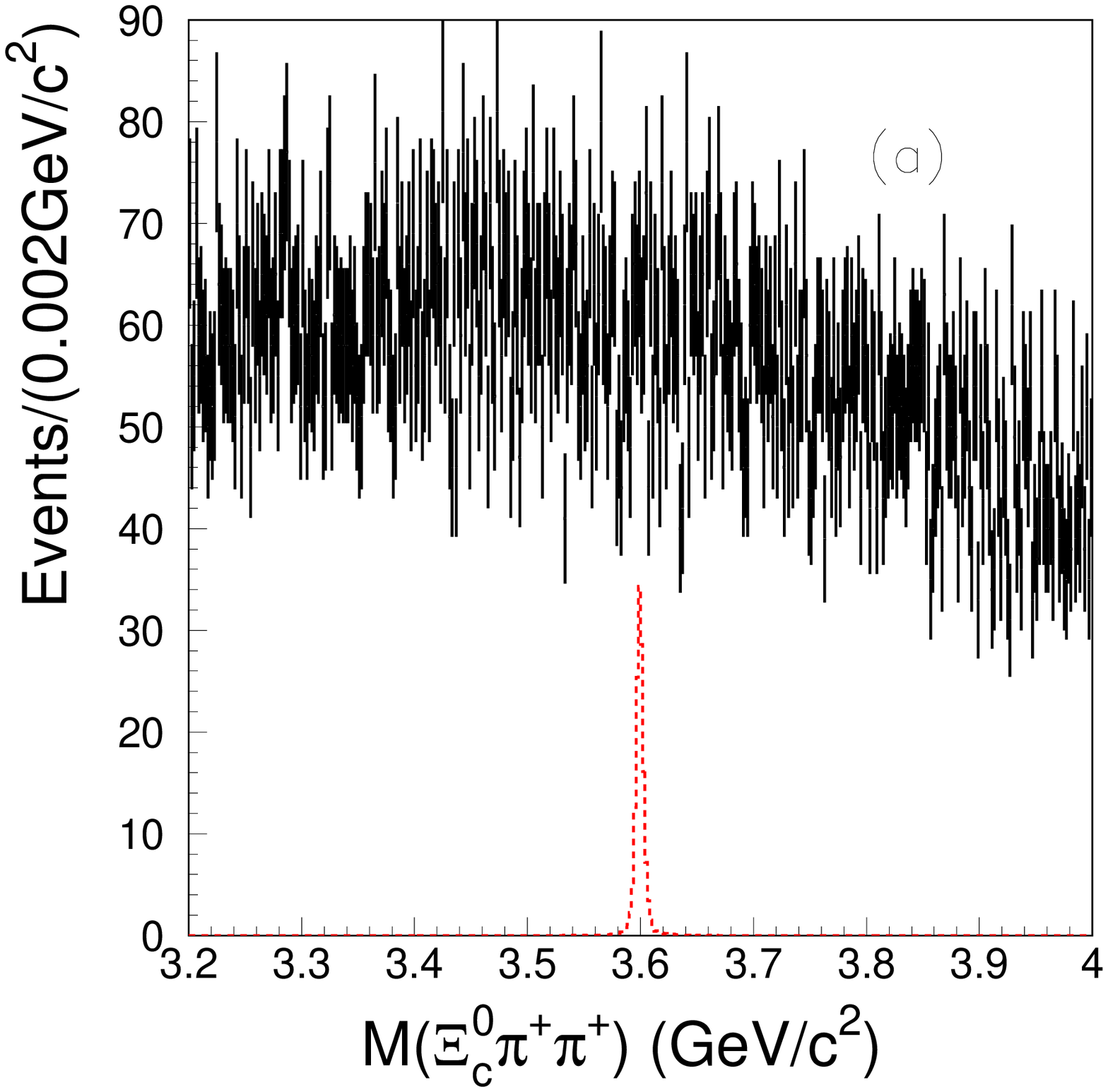}
    \includegraphics[scale=0.25]{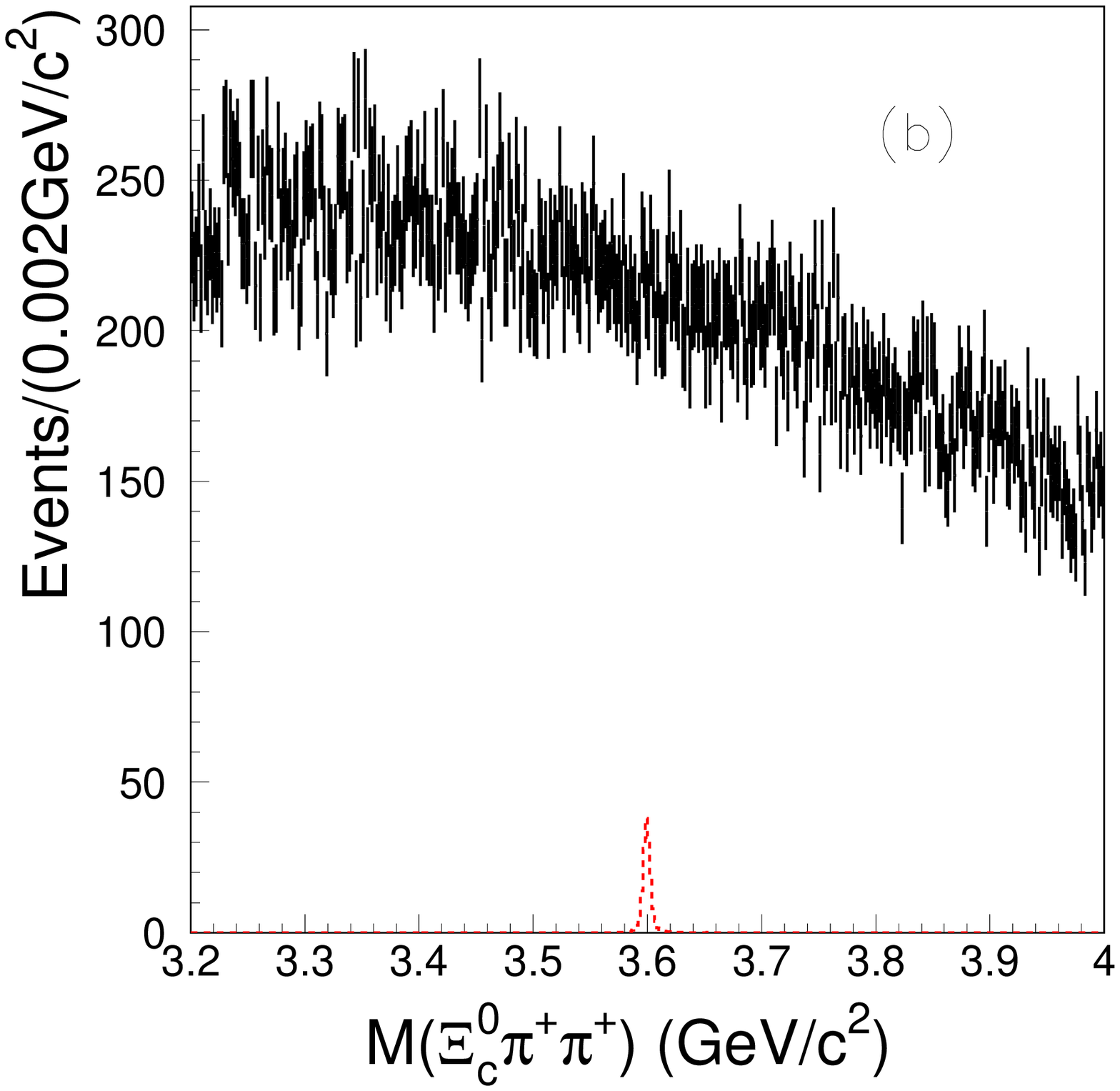}

    \includegraphics[scale=0.25]{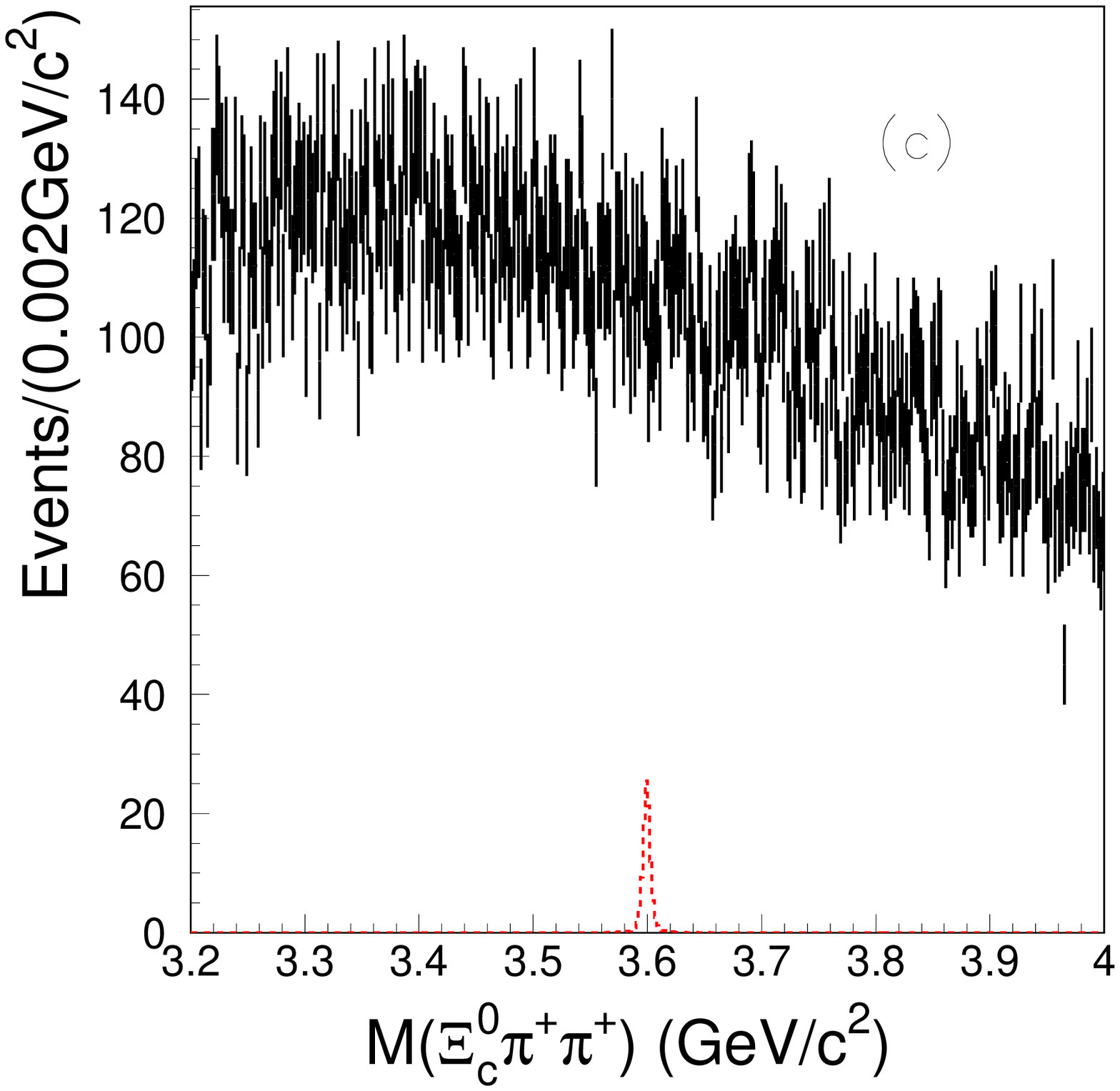}
    \includegraphics[scale=0.25]{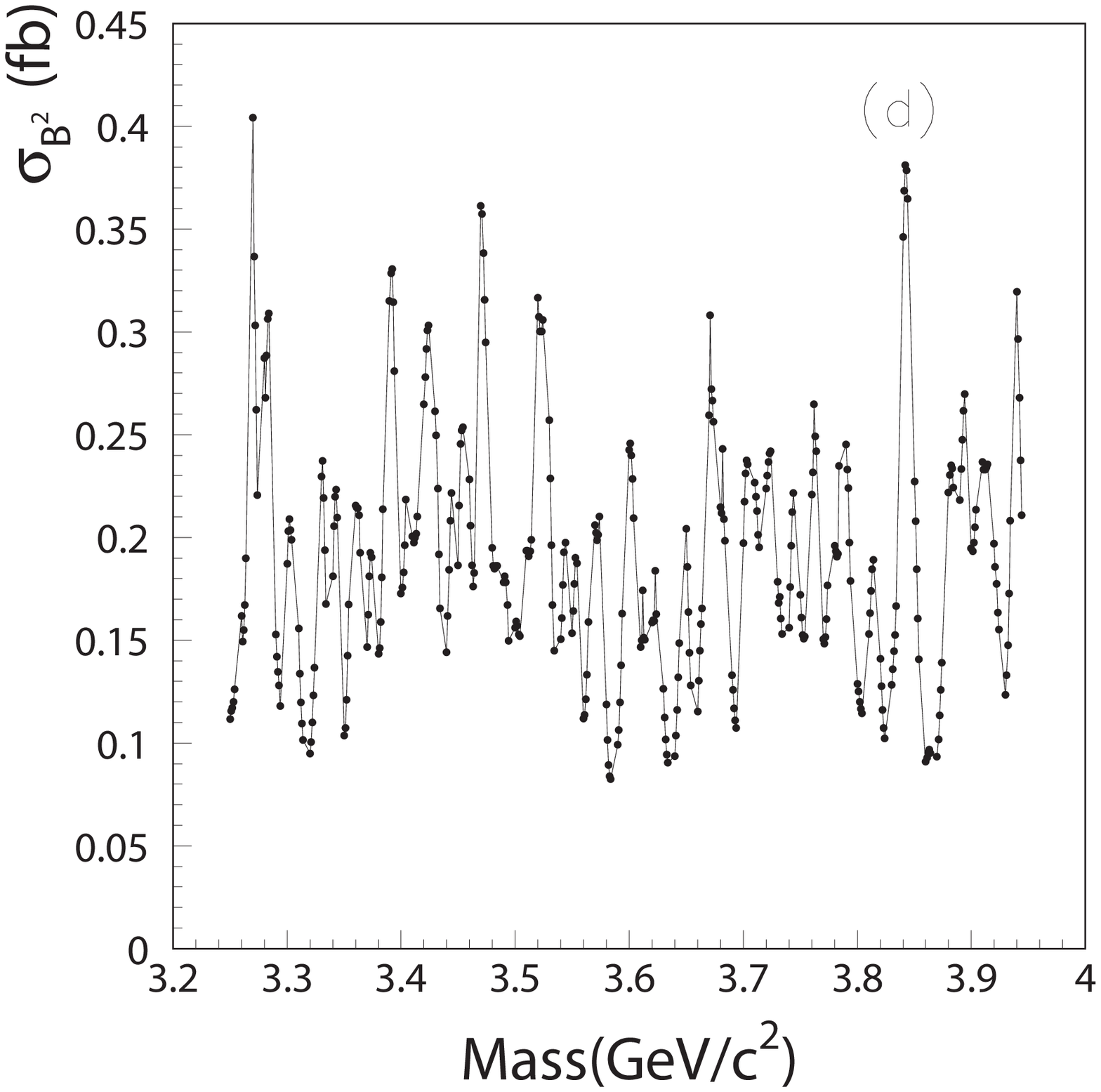}
    \caption{(a)-(c):$M(\Xi_{c}^{0}\pi^{+}\pi^{+})$ distribution in the $\Xi_{cc}^{++}$ search region for 
            $\Xi_{c}^{0}\to$  (a) $\Xi^{-}\pi^{+}$, 
            (b) $\Lambda K^{-}\pi^{+}$,
            (c) $pK^{-}K^{-}\pi^{+}$.
             The vertical error bars are from data. The dashed histograms are from signal MC.
            (d): 95$\%$ C.L. upper limit of the $\sigma_{{\cal B}^{2}}$ for $\Xi_{cc}^{++}$
                as a function of the mass with a 1 MeV/${\it c}$$^{2}$ step.}
    \label{open_guzaic+pipi_momcut}
  \end{center}
\end{figure*}

Figure \ref{open_guzaic+pi_momcut} (\ref{open_guzaic+pipi_momcut}) (a)-(c) shows the 
$M(\Xi_c^{0}\pi^{+}(\pi^{+}))$ distribution in the $\Xi_{cc}$ search region 
with all the selection 
cuts applied, overlaid with the MC expectation for the $\Xi_{cc}$ at 
the mass of 3.60 GeV/${\it c}$$^{2}$
with $\sigma(e^{+}e^{-}\to \Xi_{cc}^{+(+)} X)$ of 500 fb and 
both ${\cal B}(\Xi_{cc}^{+(+)}\to \Xi_{c}^{0}\pi^{+}(\pi^{+}))$ and
${\cal B}(\Xi_{c}^{0} \to \Xi^{-}\pi^{+})$ of 5$\%$.
The relative yields of signal MC for each decay mode are based on 
$N^{i}_{\Xi_{cc}}$.

A simultaneous UML fit, with the relative $\Xi_{cc}$ yields constrained 
as discussed earlier, 
is applied to evaluate the statistical significance of the $\Xi_{cc}$. 
The signal PDF is described with MC events generated
for each decay mode and with the $\Xi_{cc}$ mass generated in the search 
region with a 1 MeV/${\it c}$$^{2}$ step.
The mass resolution is 2.7-4.2 MeV/${\it c}$$^{2}$, depending on the mass 
of the $\Xi_{cc}$.
The background PDF is modeled as a third-order polynomial.
The highest significance is 3.2$\sigma$ for the mass 
around 3.553 GeV/${\it c}$$^{2}$ in 
the $M(\Xi_c^{0}\pi^{+})$. We perform a pseudo-experiment test 
to evaluate the probability of  observing 
a peak with such a statistical significance. 
A smooth mass distribution based on data is generated and the significance 
is evaluated in the entire search region.
The probability to observe a peak with a significance higher than
3.2$\sigma$ in one pseudo-experiment is 26$\%$. 
Therefore, the statistical significance
of 3.2$\sigma$ is insufficient to claim evidence of the $\Xi_{cc}^{+}$.

The 95$\%$ C.L. upper limit for the product of the cross section and branching fractions 
produced with $0.45<x_{p}<1.0$ condition,
\begin{eqnarray*}\label{equ_sigma2}
\sigma_{{\cal B}^{2}} &\equiv& \sigma(e^{+}e^{-}\to \Xi_{cc}^{+(+)}X) \times {\cal B}(\Xi_{cc}^{+(+)}\to \Xi_{c}^{0}\pi^{+}(\pi^{+})) \\
                      &&\times {\cal B}(\Xi_{c}^{0} \to \Xi^{-}\pi^{+})\\
&=&\frac{N_{{\rm sig}}}{2L\times \epsilon^{\Xi^{-}\pi^{+}}_{\Xi_{cc}}(1+\frac{N^{\Lambda K^{-}\pi^{+}}_{\Xi_{cc}}}{N^{\Xi^{-}\pi^{+}}_{\Xi_{cc}}}+\frac{N^{p K^{-} K^{-}\pi^{+}}_{\Xi_{cc}}}{N^{\Xi^{-}\pi^{+}}_{\Xi_{cc}}})},
\end{eqnarray*}
is evaluated with the same method as in section \ref{section_xicc_lambdac}.
In addition to the sources from the study with the $\Lambda_{c}^{+}$, 
two others are included here.
The systematic  uncertainty from the $\Lambda$ reconstruction efficiency 
is estimated to be 3$\%$ using the yield of
the $B^{+} \to \Lambda \bar{\Lambda} K^{+}$ with and without 
the requirement using decay vertex information.
The systematic  uncertainties related to $N^{i}_{\Xi_c(2645)^{+}}$ are taken 
from their statistical errors. 
The systematic  uncertainties are summarized in the fourth column in 
Table \ref{summary_sys}.
Figure \ref{open_guzaic+pi_momcut} (\ref{open_guzaic+pipi_momcut}) (d) 
shows $\sigma_{{\cal B}^{2}}$ for 
the $\Xi_{cc}^{+(+)}$ as a function of the mass with a 
1 MeV/${\it c}$$^{2}$ step.
The 95$\%$ C.L. upper limit on $\sigma_{{\cal B}^{2}}$ is 0.076--0.35 fb 
for the $\Xi_{cc}^{+}$ and 0.082--0.40 fb for the $\Xi_{cc}^{++}$.

%\begin{table*}[htbp]
%  \begin{center}
%    \caption{Summary of the systematic errors for product of cross section and branching fraction(\%)
%      ($\Xi^{-}\pi^{+}$/$\Lambda K^{-}\pi^{+}$/$pK^{-}K^{-}\pi^{+}$).}
%    \label{summary_sys_guzaicpi}
%    \begin{tabular}{|llllllll} \hline \hline
%      & pid  & Tracking & signal PDF & Luminosity & ${\cal B}$ &$N^{i}_{\Xi_c(2645)^{+}}$ \\ 
%      $\Xi_{cc}^{+}$ & 4.3/3.4/1.9 & 1.8/1.8/1.8 & 3.5  & 1.4 & 0.8/0.8/0.0 & 4.6/4.1/4.3        \\ 
%      $\Xi_{cc}^{++}$& 4.5/3.6/2.5 & 2.1/2.1/2.1 & 3.5  & 1.4 & 0.8/0.8/0.0 & 4.6/4.1/4.3        \\ \hline \hline
%    \end{tabular}
%  \end{center}
%\end{table*}

\section{Conclusion}\label{section_conclusion}
We have presented a search for doubly-charmed baryons and 
a study of the charmed strange baryons $\Xi_{c}(3055)^{+}$, 
$\Xi_{c}(3123)^{+}$ and $\Xi_{c}(2645)^{+}$
using the full data sample (980 fb$^{-1}$) collected with the Belle detector.
The search for doubly charmed baryons is an improved study of our previous 
work~\cite{Chistov:2006zj}. We use about two times statistics and 
several additional decay mode, that were not studied in the previous work.

We search for the $\Xi_{cc}$ in the $\Lambda_{c}^{+}K^{-}\pi^{+}(\pi^{+})$
and $\Xi_{c}^{0}\pi^{+}(\pi^{+})$ final states. The $\Lambda_{c}^{+}$ is 
reconstructed from
the $pK^{-}\pi^{+}$ and $pK_{S}^{0}$ decay modes.
We do not find any significant $\Xi_{cc}$ signal and set a 95$\%$ C.L.
upper limit on $\sigma(e^{+}e^{-}\to \Xi_{cc}^{+(+)} X)\times{\cal B} (\Xi_{cc}^{+(+)} \to \Lambda_{c}^{+}K^{-}\pi^{+}(\pi^{+}))$
with the scaled momentum $0.5<x_{p}<1.0$:
4.1--25.0 fb for $\Xi_{cc}^{+}$ and 2.5--26.5 fb for $\Xi_{cc}^{++}$.
We also search for the $\Xi_{cc}$ 
in the $\Xi_{c}^{0}\pi^{+}(\pi^{+})$ final state.
The $\Xi_{c}^{0}$ is reconstructed from the $\Xi^{-}\pi^{+}$,
$\Lambda K^{-}\pi^{+}$, and $pK^{-}K^{-}\pi^{+}$ decay modes. We do not 
find any significant $\Xi_{cc}$
signal and set a 95$\%$ C.L. upper limit on 
$\sigma(e^{+}e^{-}\to \Xi_{cc}^{+(+)} X) \times {\cal B} (\Xi_{cc}^{+(+)} \to \Xi_{c}^{0}\pi^{+}(\pi^{+}))
\times{\cal B} (\Xi_{c}^{0} \to \Xi^{-}\pi^{+})$ with the scaled momentum 
$0.45<x_{p}<1.0$:
0.076--0.35 fb for the $\Xi_{cc}^{+}$ and 0.082--0.40 fb for the $\Xi_{cc}^{++}$.
When we compare these values with the measurements by BaBar, we should note several things.
We should multiply our result by two as written in section \ref{section_lambdac}. 
For the final states with $\Lambda_{c}^{+}$, their upper limit is for the product of cross section, 
${\cal B} (\Xi_{cc}^{+(+)} \to \Lambda_{c}^{+}K^{-}\pi^{+}(\pi^{+}))$ and ${\cal B} (\Lambda_{c}^{+}\to p K^{-}\pi^{+})$.
The values presented in Ref. \cite{Aubert:2006qw} are the highest upper limits in the search region,
which can be compared with our highest values. After taking into account these points, we find
our limits represent improvements by about a factor two for the final states with $\Lambda_{c}^{+}$ 
and a factor of four for the final states with $\Xi_{c}^{0}$.

If we assume ${\cal B} (\Xi_{cc}^{+(+)} \to \Lambda_{c}^{+}K^{-}\pi^{+}(\pi^{+}))$, ${\cal B} (\Xi_{cc}^{+(+)} \to \Xi_{c}^{0}\pi^{+}(\pi^{+}))$
and ${\cal B} (\Xi_{c}^{0} \to \Xi^{-}\pi^{+})$ to be 5$\%$, 
which is equal to the  ${\cal B} (\Lambda_{c}^{+}\to p K^{-} \pi^{+})$, 
the upper limits on the $\sigma(e^{+}e^{-}\to \Xi_{cc}^{+}X)$ are
82--500 fb ($\Xi_{cc}^{+}$) and 50--530 fb ($\Xi_{cc}^{++}$)
for the decay mode with the $\Lambda_{c}^{+}$ and 30--140 fb 
($\Xi_{cc}^{+}$) and 33--160 fb ($\Xi_{cc}^{++}$) for the 
decay mode with the $\Xi_{c}^{0}$. These values are comparable to some
of the theoretical predictions~\citep{Kiselev:1994pu,Ma:2003zk}.

We have searched for the $\Xi_{c}(3055)^{+}$ and $\Xi_{c}(3123)^{+}$ 
in the $\Lambda_{c}^{+}K^{-}\pi^{+}$ decays through intermediate
$\Sigma_{c}(2455)^{++}$ or $\Sigma_{c}(2520)^{++}$ states. 
We observe the $\Xi_{c}(3055)^{+}$ with a significance of
6.6$\sigma$, including systematic uncertainty.
The mass and width are measured to be 
3058.1 $\pm$ 1.0 (stat) $\pm$ 2.1 (sys) MeV/${\it c}$$^{2}$ and 
9.7 $\pm$ 3.4 (stat) $\pm$ 3.3 (sys) MeV/${\it c}$$^{2}$, respectively.
We do not observe any significant signal corresponding to 
the $\Xi_{c}(3123)^{+}$.
%and 95$\%$ C.L.
%upper limit on the $\sigma(e^{+}e^{-}\to \Xi_{c}(3123)^{+} X)\times {\cal B}(\Lambda_{c}^{+} \to pK^{-}\pi^{+})$,
%with the scaled momentum $x_{p}>0.7$, is measured to be 0.17 fb,
%which is much smaller than BaBar's measurement(1.6 $\pm$ 0.6 $\pm$ 0.2 fb).

The first measurement of the width of the $\Xi_{c}(2645)^{+}$ has been 
also performed, yielding 2.6 $\pm$ 0.2 (stat)$\pm$ 0.4 (sys) MeV/${\it c}$$^{2}$.
%The measured width can give a constraint on the theoretical models.

\acknowledgments

We thank the KEKB group for the excellent operation of the
accelerator; the KEK cryogenics group for the efficient
operation of the solenoid; and the KEK computer group,
the National Institute of Informatics, and the 
PNNL/EMSL computing group for valuable computing
and SINET4 network support.  We acknowledge support from
the Ministry of Education, Culture, Sports, Science, and
Technology (MEXT) of Japan, the Japan Society for the 
Promotion of Science (JSPS), and the Tau-Lepton Physics 
Research Center of Nagoya University; 
the Australian Research Council and the Australian 
Department of Industry, Innovation, Science and Research;
Austrian Science Fund under Grant No. P 22742-N16;
the National Natural Science Foundation of China under contract 
No.~10575109, 10775142, 10825524, 10875115, 10935008 and 11175187; 
the Ministry of Education, Youth and Sports of the Czech 
Republic under contract No.~MSM0021620859;
the Carl Zeiss Foundation, the Deutsche Forschungsgemeinschaft
and the VolkswagenStiftung;
the Department of Science and Technology of India; 
the Istituto Nazionale di Fisica Nucleare of Italy; 
The WCU program of the Ministry Education Science and
Technology, National Research Foundation of Korea Grant No.
2011-0029457, 2012-0008143, 2012R1A1A2008330, 2013R1A1A3007772,
BRL program under NRF Grant No. KRF-2011-0020333, BK21 Plus program,
and GSDC of the Korea Institute of Science and Technology Information;
the Polish Ministry of Science and Higher Education and 
the National Science Center;
the Ministry of Education and Science of the Russian
Federation and the Russian Federal Agency for Atomic Energy;
the Slovenian Research Agency;
the Basque Foundation for Science (IKERBASQUE) and the UPV/EHU under 
program UFI 11/55;
the Swiss National Science Foundation; the National Science Council
and the Ministry of Education of Taiwan; and the U.S.\
Department of Energy and the National Science Foundation.
This work is supported by a Grant-in-Aid from MEXT for 
Science Research in a Priority Area (``New Development of 
Flavor Physics''), from JSPS for Creative Scientific 
Research (``Evolution of Tau-lepton Physics''), and 
Grant-in-Aid for Scientific Research on Innovative Areas ''Elucidation of 
New Hadrons with a Variety of Flavors''.


\begin{thebibliography}{99}
\bibitem{CC}
Throughout this paper, 
the inclusion of the charge-conjugate decay mode is implied
unless otherwise stated.
\end{thebibliography}
\end{document}